\documentclass{aa}
\usepackage{graphicx}
\usepackage{amssymb}
\usepackage{natbib}
\bibpunct{(}{)}{;}{a}{}{,}

\begin{document}

\title{What is the temperature structure in the giant H{\sc ii} region
\object{NGC 588}?}

\titlerunning{The giant H{\sc ii} region \object{NGC 588}}

\author{L. Jamet\inst{1,2} \and G. Stasi\'nska\inst{1} \and E. P\'erez\inst{2} 
\and R.M. Gonz\'alez Delgado\inst{2} \and J.M. V{\'\i}lchez\inst{2}}

\offprints{L. Jamet, \email{luc@iaa.es}}

\institute{LUTH, Observatoire de Meudon, 5 place Jules Janssen, 92195 Meudon
Cedex, France \and Instituto de Astrof{\'\i}sica de Andaluc{\'\i}a (CSIC),
Apartado 3004, 18080 Granada, Spain}


\date{Received / Accepted }

\abstract{
We present the results of an exhaustive study of the ionized gas in \object{NGC
588}, a giant H{\sc ii} region in the nearby spiral galaxy \object{M33}. This
analysis uses a high number of diagnostics in the optical and infrared ranges.
Four temperature diagnostics obtained with optical lines agree with a gas
temperature of 11000K, while the [O{\sc iii}] $\lambda$5007/$\lambda$88$\mu$m
ratio yields a much lower temperature of $\approx$8000K. This discrepancy
suggests the presence of large temperature inhomogeneities in the nebula. We
investigated the cause of this discrepancy by constructing photoionization
models of increasing complexity. In particular, we used the constraints from
the H$\alpha$ and H$\beta$ surface brightness distributions and
state-of-the-art models of the stellar ionizing spectrum. None of the
successive attempts was able to reproduce the discrepancy between the
temperature diagnostics, so the thermal balance of \object{NGC 588} remains
unexplained. We give an estimate of the effect of this failure on the O/H and
Ne/O estimates and show that O/H is known to within $\pm$0.2 dex.

}

\maketitle

\section{Introduction}
Giant H{\sc ii} regions (GHRs) are the most popular tracers of elemental
abundances in distant galaxies, since they are easy to observe and, in
principle, easy to analyze. Abundances can be obtained directly from the
intensities of emission lines, without having to go through a detailed model
analysis to extract the information \citep[e.g.,][]{gst04}. However, for
metal-rich H{\sc ii} regions -- say, with an oxygen abundance larger than half
that of the Sun -- the methods for abundance derivation are statistical and
rely on previous calibrations based on photoionization models. The large number
of calibrations that have been published, leading to significantly different
abundances \citep[e.g.,][]{kbg03}, show that the interpretation of emission
lines in H{\sc ii} regions may not be as easy as one would like. In the case of
metallicities significantly below solar, abundance determination is considered
more reliable, since the optical emission lines provide a direct measure of the
gas temperature, which allows one to compute the elemental abundances directly
from the observed line intensities. However, as shown by \cite{p67}, if the
temperature in H{\sc ii} regions is not uniform but instead presents spatial
fluctuations, the derived abundances will be biased. The reality of the
presence of these temperature fluctuations and their cause are still a matter
of debate \citep[e.g.,][]{e02}.

Comprehensive analysis of nearby giant H{\sc ii} regions are necessary to check
our understanding of the thermal structure of these objects and to validate
empirical methods for abundance determinations. There have already been
detailed photoionization studies of some giant H{\sc ii} regions
\citep{gvg97,ss99,lpl99,lpp03,gdp00,lp01,rpb02}. In most cases, the models were
not able to reproduce the observed temperature indicators correctly. However,
this was not the main goal of most of those studies, and the degree of
sophistication of the models was perhaps insufficient for that purpose.

In the present paper, we propose a comprehensive analysis of \object{NGC 588},
a giant H{\sc ii} region located on the outskirts of the nearby spiral galaxy
\object{M33}, with the aim of understanding its temperature structure. We
gathered a large set of spectroscopic and imaging data in various wavelength
ranges. From this set of data, we were able to give a full description of the
ionizing stellar population, based on a star-by-star analysis \citep{jpc04}. We
now use the ionizing radiation field from this population together with
information on the nebular morphology given by narrow band images to construct
photoionization models. Constraints are provided by the strengths of optical
and infrared emission lines. The availability of infrared data is particularly
important, since they enlarge the number of possible spectral diagnostics.

The paper is organized as follows. In Sect. \ref{sect_obsred} we describe the
observational data and their processing. In Sect. \ref{sect_empir} we present
empirical diagnostics of the density, temperature, and chemical composition. In
Sect. \ref{sect_mdl} we give details on the model-fitting strategy that we
adopted. Starting from very simple models (Sect. \ref{sect_simple}), we
gradually increased the degree of sophistication in order to match the
observations as closely as possible (Sect.
\ref{sect_hbo2o3}--\ref{sect_further}). We then examine the effects of energy
sources other than the ionizing flux of the cluster (Sect. \ref{sect_other}).
In Sect. \ref{sect_ohneo}, we evaluate the impact of the unknowns of the
thermal structure of the nebula on the determination of the O/H and Ne/O
abundance ratios. Finally, we present our conclusions in Sect.
\ref{sect_concl}.
\section{Observational data and reduction}
\label{sect_obsred}
\begin{figure}
\resizebox{\hsize}{!}{\includegraphics{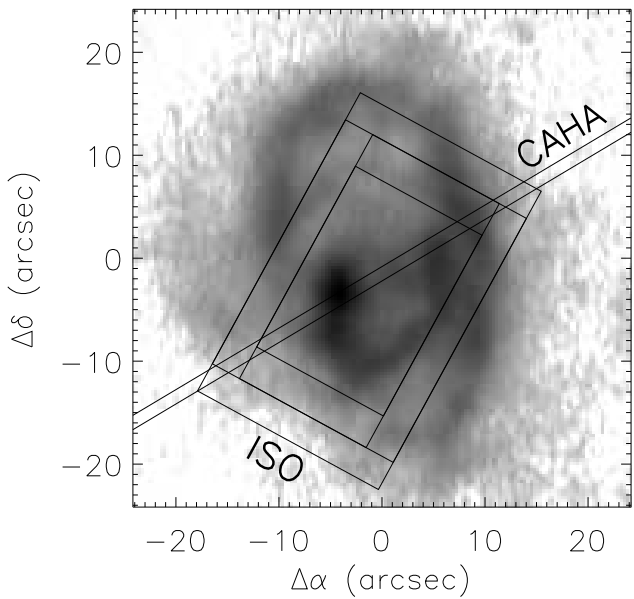}}
\caption{H$\alpha$ image with the spectroscopic slits overplotted. North is up,
and east is on the left. The image display scale is logarithmic.}
\label{fig_ha_slits}
\end{figure}

We acquired ground-based optical spectra of \object{NGC 588} and retrieved a
series of data available for this object in the archives of the Infrared Space
Observatory (ISO) and of the Isaac Newton Group (ING). The main properties of
the data with which we realized measurements are summarized in Table
\ref{tab_obs}. Figure \ref{fig_ha_slits} shows the nebula in H$\alpha$ (see
Sect. \ref{sect_ing}) on which the spectroscopic slits were superimposed.

\begin{table}
\caption{Journal of Observations.}
\label{tab_obs}
\center
\begin{tabular}{ccc}
\hline
\hline
\multicolumn{3}{l}{\bf Spectroscopy} \\
\hline
\multicolumn{3}{l}{CAHA (PA=121$^{\circ}$, width=1.2$\arcsec$, 1998-08-27/28)}\\
\hline
range: $\lambda$ ({\AA}) & $\Delta\lambda$ ({\AA}/pixel) & exposure (s) \\
B : 3595--5225 & 0.81 & $4\times1800$ \\
R1: 5505--7695 & 1.08 & $2\times1800$ \\
R2: 7605--9805 & 1.08 & $2\times1800$ \\
\hline
\multicolumn{3}{l}{ISO/SWS (PA=$-29^{\circ}$, data set 81601775, 1998-02-08)} \\
\hline
line & range ($\mu$m) & aperture ($''$) \\
{}[S{\sc iv}] $\lambda$11$\mu$m   & 10.45--10.55 & 14$\times$20 \\
{}[Ne{\sc iii}] $\lambda$16$\mu$m & 15.46--15.64 & 14$\times$27 \\
{}[S{\sc iii}] $\lambda$19$\mu$m  & 18.61--18.77 & 14$\times$27 \\
{}[O{\sc iv}] $\lambda$26$\mu$m   & 25.80--26.08 & 14$\times$27 \\
{}[S{\sc iii}] $\lambda$34$\mu$m  & 33.17--33.75 & 20$\times$33 \\
{}[Ne{\sc iii}] $\lambda$36$\mu$m & 35.70--36.26 & 20$\times$33 \\
\hline
\multicolumn{3}{l}{ISO/LWS} \\
\hline
Data set & Spectrometer & Exposure (s) \\
59901081 & LWS01 & 1478 \\
80800268 & LWS02 & 810 \\
81601776 & LWS01 & 1402 \\
\hline
\hline
\multicolumn{3}{l}{\bf Imaging} \\
\hline
\multicolumn{3}{l}{JKT (1990-08-20/21)} \\
\hline
Band      & Filter & Exposure (s)      \\
H$\beta$  & 4861/54 & $2\times1800$    \\
H$\alpha$ & 6563/53 & 1800             \\
H$\alpha$ continuum & 6834/51 & 1300   \\
{}[O{\sc iii}] $\lambda$5007 & 5007/51 & 1800   \\
\hline
\end{tabular}
\end{table}

\subsection{Optical CAHA spectra}
We obtained a set of 3 spectra on 27 August 1998 with the 3.5m telescope of the
Centro Astron\'omico Hispano Alem\'an (CAHA, Calar Alto, Almer{\'\i}a, Spain).
These observations were acquired with the dual beam TWIN spectrograph, equipped
with a beam splitter centered on 5500 {\AA} which sends the signal through two
arms (``red'' and ``blue''). The spectrum observed with each arm was collected
with an SITe-CCD of 2000$\times$800 15$\mu$m pixels. We took four exposures in
the range 3595--5225 {\AA} (range B) with the blue arm, two in the range
5505--7695 {\AA} (range R1), and two in the range 7605--9805 {\AA} (range R2)
with the red arm; each of these exposures was 1800 s long. The two gratings
used for the observations, T7 (blue arm) and T4 (red arm), give dispersions of
0.81 and 1.08 {\AA}/pix, respectively. The signal was acquired through a
$240\times1.2''$ slit oriented at PA=121$^\circ$ with a resolution of $\sim$2
{\AA} along the total spectral range, as measured from the FWHM of the lines of
the sky background.

\subsubsection{Reduction}
\label{sect_cahareduc}
We reduced the bidimensional spectra following the standard procedure, mainly
making use of the IRAF\footnote{IRAF is distributed by the National Optical
Astronomy Observatory, which is operated by the Association of Universities for
Research in Astronomy, Inc., under cooperative agreement with the National
Science Foundation.} package {\tt noao.twodspec.longslit}.

The flat-field correction was made with two light sources: an internal tungsten
lamp for removal of the effect of pixel-to-pixel sensitivity variations and the
twilight sky for correction for instrumental vignetting. Though the tungsten
lamp spectra were heavily affected by fringing effects, the overal sensitivity
flattening was satisfactory; we estimated the residuals to be $\sim$1\%.

The wavelength calibration was performed from spectra of a helium-argon arc
lamp, typicall givingy 20 useful lines per frame. Several of these spectra were
obtained throughout the night in order to account for instrumental drifts.

No data was available to calibrate the frames with respect to the positions
along the slit, and the inspection of several spectra of stars showed $\sim$2
pixel ($\approx 1''$) distorsions around horizontal lines of pixels. Since we
were mainly interested in integrated line fluxes, we ignored this issue, except
for the hydrogen lines used for dereddening (see Sect. \ref{sect_hdered}).

We carefully combined the individual exposures in each range (B, R1, R2). We
first performed slight shifts of these frames along the slit axis, to match the
profiles of the lines, since the instrument centering on the object was subject
to small variations between the different exposures. The differences between
the coregistered frames showed no detectable residuals other than cosmic rays,
except for some of the brightest lines (H$\alpha$, H$\beta$, and [O{\sc iii}]
$\lambda$5007) in a $\sim$4$''$ zone around the maximum nebular emission.
However, these discrepancies were small and resulted in flux errors less than
$\sim$3\% in this zone. Since this error is small and concerns only a small
zone of the slit, we neglected it.

The combined frames were then flux-calibrated. To this aim, the photometric
response of the instrument was measured with three standard stars,
\object{BD+28~4211}, \object{G191-B2B} and \object{GD71}, observed with a
3.6$''$ wide slit and through airmasses similar to the one of the observations
of \object{NGC 588} (less than 1.2). We excluded the stellar lines from the
measurement points, because of the difference of spectral resolution between
our observations and the reference spectra. The atmospheric extinction function
we adopted is the average one of the Observatorio de Roque de los Muchachos (La
Palma, Spain), located at an altitude very similar to the one at CAHA. In each
of the three spectral ranges, we found the response measures, as obtained with
the different available spectra, to show the same chromatic trends, but
constant discrepancies of $\sim$0.1 magnitude. We also found oscillations of
amplitude $\approx$3\% on scales of 50--200 {\AA}, which we were unable to fit.
Because of the drop in the instrumental response, the $\lambda<4000$ {\AA}
range is more uncertain at $\sim$15\% with respect to the overall range B.
Furthermore, in our data, a series of wide regions in the range R1 and, above
all, R2 suffer severe extinction by telluric O$_2$ and H$_2$O unresolved
molecular bands. Such extinction tended to cause numerical instabilities when
fitting smooth functions on the measured photometric response, so we decided to
correct them, even coarsely, to avoid these instabilities. For this, we
synthesized the O$_2$ and H$_2$O molecular bands, starting from individual
lines as catalogued in the solar atlases of \cite{mmh66} and \cite{m50}. In
both atlases, only wavelength centers $\lambda_i$ and line strengths $W_i$ are
available, and we decided to model the bands as functions of magnitude loss of
the form $F(\lambda;{\rm X})=N({\rm X})\sum_i W_i H(\lambda-\lambda_i)$, with
${\rm X}$=O$_2$ or H$_2$O, $N({\rm X})$, the relative column density of ${\rm
X}$ on the line of sight, and $H({\lambda})$, the Gaussian spectral PSF of the
spectrograph. We then corrected the observed spectra of the standard stars for
these extinction functions, using values of $N({\rm O_2})$ and $N({\rm H_2O})$
that smoothed the corrected spectra best, and we computed the instrumental
response in the ranges R1 and R2 with them.

We checked the photometric consistency between the computed responses in the
three ranges B, R1, and R2. For this, we calibrated the three corresponding
spectra of one of the standard source, \object{BD+28 4211}, observed in the
same time interval, and searched for possible discontinuities from range to
range. We found no such discontinuity, meaning that the three computed
responses are mutually consistent from the point of view of absolute
photometry.

\begin{figure*}
\resizebox{\hsize}{!}{\includegraphics{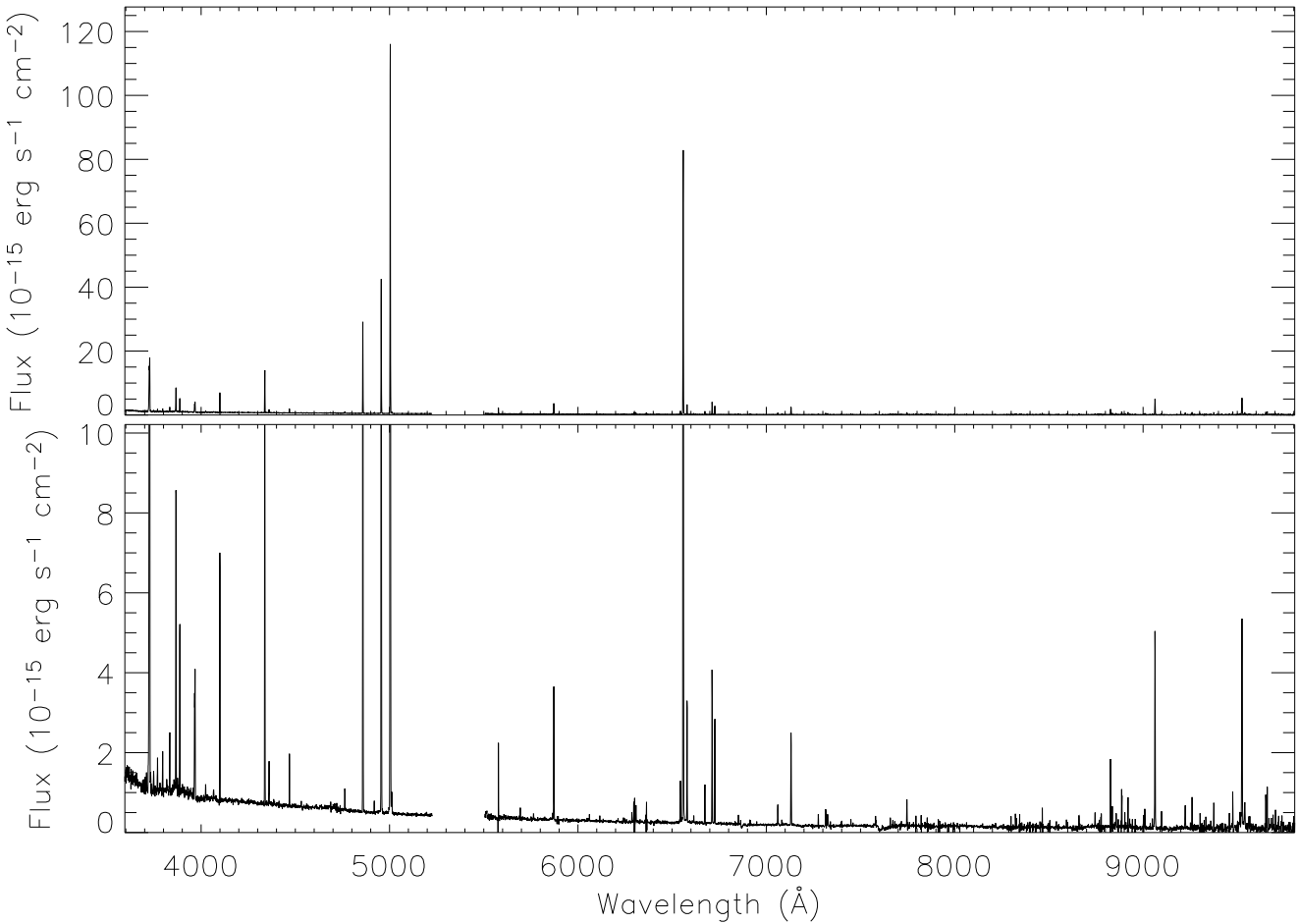}}
\caption{Optical spectrum of \object{NGC 588} shown in two flux scales.}
\label{fig_spect}
\end{figure*}

The last step in the calibration of the optical spectra was sky background
removal. At each wavelength, this background was taken as a linear ramp fitted
on two zones of the slit situated on both sides of the nebula and free of
nebular or stellar emission. Figure \ref{fig_spect} shows the optical spectrum
along the whole wavelength range covered, integrated over the 81-pixel ($45''$)
zone of the slit where the extinction was computed.

\subsubsection{Estimation of the reddening and of the dereddened H$\beta$ flux}
\label{sect_hdered}
Once the spectra were reduced, we first performed an estimation of the profiles
of the color excess $E_{B-V}(x)$ and of the dereddened H$\beta$ flux
$F_{\beta}^{\rm der}(x)$ along the slit, using the case-B line ratios of the
hydrogen Balmer decrement and assuming a gas temperature of 11000 K
\citep{vpd88}. We compared the fluxes of the Balmer lines H$\alpha$, H$\beta$,
H$\gamma$, and H$\delta$ at the different positions $x$ along the slit common
to both ranges B and R1. The idea underlying the use of these four lines
instead of two is the possibility of correcting the evaluations of $E_{B-V}(x)$
and $F_{\beta}^{\rm der}(x)$ for the errors in the line fluxes. Such errors
were split in three components: photometric residuals, underlying stellar
absorption lines, and data noise.

We first measured the fluxes $F_X(x)$ ($X=\alpha$, $\beta$, $\gamma$,
$\delta$). For each line and at each position, the flux was measured as the sum
of the pixels covered by the line, from which the continuum was previously
removed. The fluxes were measured on spectral windows wide enough to include
the underlying stellar absorption lines entirely. Then, we applied slight
shifts to the line flux profiles, in order to account for object distortions in
the frames (see Sect. \ref{sect_cahareduc}). We also convolved the H$\alpha$
profile by a mask so that its PSF along the slit matched the one of the three
other lines. Finally, we dereddened the fluxes from foreground Galactic
reddening, using the Galactic extinction law \citep{nth75,s79} and
$E_{B-V}=0.045$ \citep{bh84}.

\begin{figure}
\resizebox{\hsize}{!}{\includegraphics{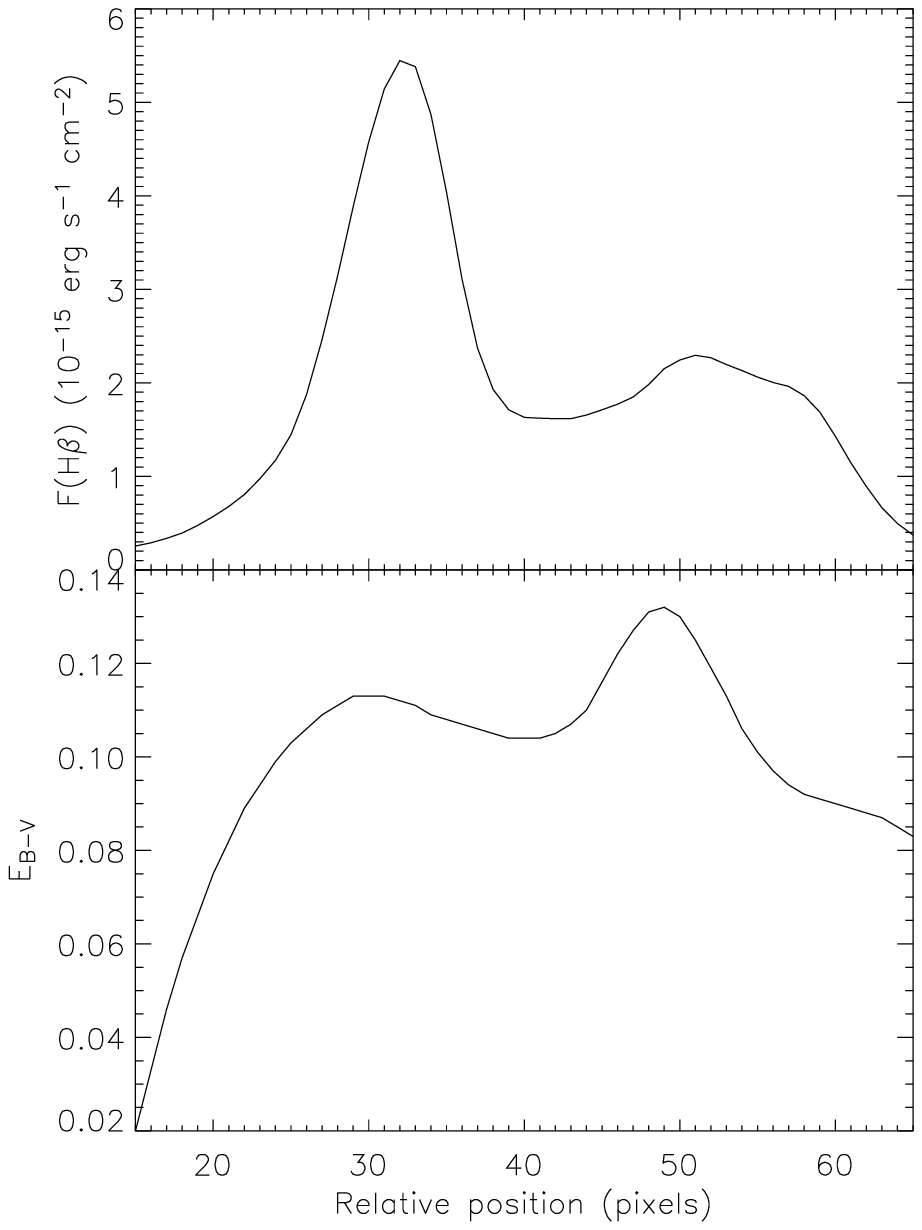}}
\caption{Curves of $E_{B-V}(x)$ and $F_{\beta}^{\rm der}(x)$ along the CAHA
slit. The $E_{B-V}(x)$ curve does not include the foreground Galactic
contribution.}
\label{fig_ebv_fhb}
\end{figure}
\begin{figure}
\resizebox{\hsize}{!}{\includegraphics{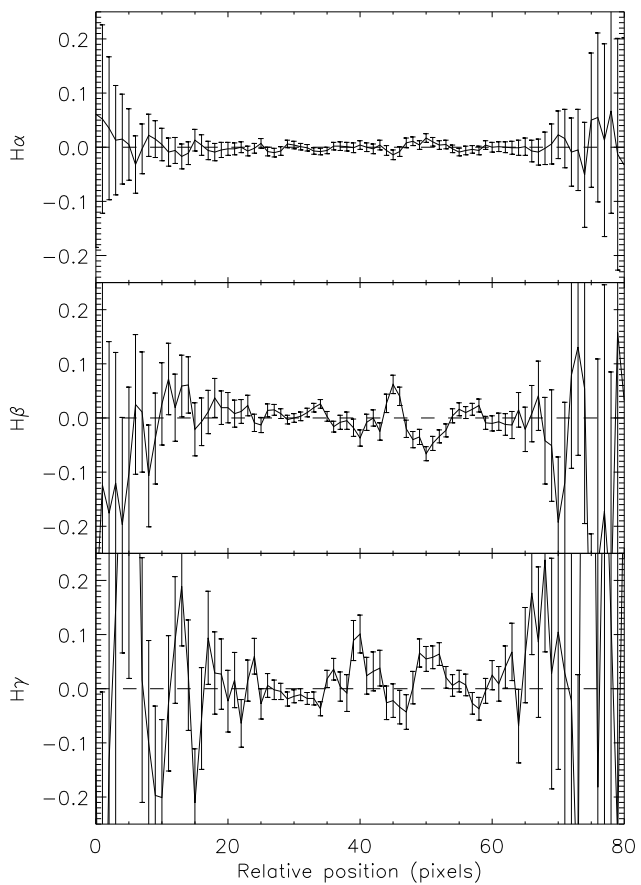}}
\caption{Residuals of the computation of $F_{\beta}^{\rm der}(x)$ and
$E_{B-V}(x)$.}
\label{fig_extresid}
\end{figure}

Using the LMC law \citep{h83} for the extinction due to M33, we then proceeded
to correction of the line fluxes for photometric residuals and for underlying
stellar absorption lines, using the procedure explained in Appendix
\ref{app_dered}. Finally, from the corrected profiles $F_{X}^{\rm corr}(x)$
($X=\alpha$, $\beta$, $\gamma$), we proceeded to the fit of $E_{B-V}(x)$ and
$F_{\beta}^{\rm der}(x)$. The $E_{B-V}(x)$ curve was smoothed with a Gaussian
pattern of 6 pixels FWHM, because of significant noise in some zones of the
slit. On the edges of the nebula, the smoothed curve was slightly negative so
we clipped it to 0. Both $E_{B-V}(x)$ and $F_{\beta}^{\rm der}(x)$ curves are
shown in Fig. \ref{fig_ebv_fhb}. As an indication of the quality of the
dereddening procedure, Fig. \ref{fig_extresid} shows the relative residuals
$F_{X}^{\rm corr}(x)/F_{X}^{\rm pred}(x)-1$, where $F_{X}^{\rm pred}(x)$ is the
flux of H$X$ predicted from $F_{\beta}^{\rm der}(x)$, $E_{B-V}(x)$ and the
theoretical H$X$/H$\beta$ ratio, for the H$\alpha$, H$\beta$, and H$\gamma$
lines. Unexplained systematic residuals of a few \% are found for H$\beta$ and
H$\gamma$ around $x=250$; but by changing artificially the line fluxes, we
found that in this part of the slit, the uncertainty on $F_{\beta}^{\rm
der}(x)$ and $E_{B-V}(x)$ is increased by less than 2\% and 0.01, respectively.

\subsubsection{Line listing and measurement}
We established a list of detected lines from visual inspection of the
calibrated frames and selected the ones to be measured. Each line was fitted by
a Gaussian curve in a series of bins along the slit. The weakest lines were
fitted in bins of several pixels whose sizes were ruled by their overall
signal-to-noise ratios (S/N), so as to ensure the convergence of the Gaussian
curve fits. The fits were performed with the program TWODSPEC of the
STARLINK\footnote{The STARLINK Project is run by CCLRC on behalf of PPARC.
TWODSPEC was developed by T.N. Wilkins and D.J. Axon.} software suite; the
[O{\sc ii}] $\lambda$3726+3729 doublet and of the [O{\sc i}] $\lambda$6300
line, blended with an emission line of the sky, were measured with a personal
IDL routine. Each bin was then dereddened using the corresponding value of
$E_{B-V}$. Finally, to obtain the line intensities with respect to H$\beta$, we
summed the dereddened line fluxes in all the bins and divided the result by the
dereddened H$\beta$ flux summed over the same bins. In this procedure, we
included the bins where $E_{B-V}$ was clipped to 0.

\subsubsection{Detection limit and accuracy of the measured fluxes}
From measurements of the noise in the frames, we estimated the 3$\sigma$
detection limit of emission lines in the spectra. We found this limit to vary
approximately from $F({\rm H}\beta)/2000$ ($\lambda\geq 5000$ {\AA}) to $F({\rm
H}\beta)/500$ ($\lambda\approx 3700$ {\AA}). The S/N was then very high for the
main lines. However, recombination lines of metals remained undetectable. In
particular, the 3$\sigma$ detection threshold for the C{\sc ii} $\lambda$4267
and O{\sc ii} $\lambda$4651 was $\approx F({\rm H}\beta)/500$.

As already mentioned, in each spectral range, we found oscillations of
$\sim$3\% amplitude in the residuals of the fit of the photometric instrumental
response. These oscillations are even higher below $\lambda=4000$ {\AA},
causing a $\approx15$\% uncertainty in the photometric match of this range with
the rest of the B range. Hence, when calculating line intensity ratios, we
decided to add 3\% of uncertainty in the fluxes of the lines with
$\lambda>4000$ {\AA}, and 15\% for lines with $\lambda\leq 4000$ {\AA}. There
are two exceptions to this rule: the H$\beta$ line (see Sect.
\ref{sect_hdered}) and the ratios of lines belonging to a close doublet and
then suffering identical photometric errors (e.g., [O{\sc ii}]
$\lambda$3726/$\lambda$3729, [S{\sc ii}] $\lambda$6731/$\lambda$6716).

Telluric absorption bands are another important source of uncertainty. The way
we corrected the calibration spectra for their effect is very approximate and
let in an uncertainty of $\sim$30\% at $\lambda>8200$ {\AA}. However, the main
source of error in this range comes from the lack of knowledge about the
coincidence of the individual (unresolved) lines of the bands with the emission
lines of the nebula, especially [S{\sc iii}] $\lambda$9069 and [S{\sc iii}]
$\lambda$9532. Indeed, the flux of a nebular line can be severely reduced,
depending on its exact position with respect to the telluric absorption lines.
Consequently, we did not use the fluxes of the lines situated at $\lambda>8200$
{\AA}.

Finally, an additional uncertainty in each line ratio results from a constant
uncertainty of 0.02 in the $E_{B-V}$ curve. Table \ref{tab_intens} presents the
reddening-corrected intensities of all the lines measured in the spectrum and
used in this work, in units of ${\rm H}\beta=100$, and the associated standard
deviations (i.e. the uncertainties related to random errors only).

\begin{table}
\caption{Dereddened fluxes of the lines and doublets used in this work, in
units of H$\beta=100$, and associated measure standard deviations in the same
units.$^a$Fluxes corrected for the SWS apertures.}
\label{tab_intens}
\center
\begin{tabular}{lll}
\hline
\hline
Line/Doublet & Flux & Standard deviation \\
\hline
\hline
Optical slit \\
\hline
{}[O{\sc ii}] $\lambda$372(6+9)   & 141.54 & 0.36 \\
{}[Ne{\sc iii}] $\lambda$3869     & 36.86  & 0.18 \\
{}[S{\sc iii}] $\lambda$40(69+76) & 1.58   & 0.14 \\
{}[O{\sc iii}] $\lambda$4363      & 3.67   & 0.07 \\
{}[Ar{\sc iv}] $\lambda$4711      & 0.26   & 0.07 \\
{}[Ar{\sc iv}] $\lambda$4741      & 0.34   & 0.05 \\
{}[O{\sc iii}] $\lambda$5007      & 404.79 & 0.20 \\
{}[N{\sc ii}] $\lambda$5755       & 0.20   & 0.03 \\
He{\sc i} $\lambda$5876           & 11.94  & 0.04 \\
{}[S{\sc iii}] $\lambda$6312      & 1.38   & 0.04 \\
{}[N{\sc ii}] $\lambda$6583       & 10.69  & 0.04 \\
{}[S{\sc ii}] $\lambda$6716       & 12.73  & 0.04 \\
{}[S{\sc ii}] $\lambda$6731       & 9.17   & 0.03 \\
{}[Ar{\sc iii}] $\lambda$7135     & 9.47   & 0.03 \\
{}[O{\sc ii}] $\lambda$73(20+30)  & 3.09   & 0.05 \\
\hline
\hline
Integrated fluxes \\
\hline
{}[O{\sc iii}] $\lambda$5007          & 352.22 & 12.46 \\
{}[S{\sc iv}] $\lambda$11$\mu$m$^a$   & 37.81  & 9.72 \\
{}[Ne{\sc iii}] $\lambda$16$\mu$m$^a$ & 56.51  & 4.75 \\
{}[S{\sc iii}] $\lambda$19$\mu$m$^a$  & 81.42  & 13.66 \\
{}[S{\sc iii}] $\lambda$33$\mu$m$^a$  & 56.23  & 9.73 \\
{}[O{\sc iii}] $\lambda$88$\mu$m      & 214.47 & 8.18 \\
\end{tabular}
\end{table}

\subsection{Ground-based images}
\label{sect_ing}
\begin{figure}
\resizebox{\hsize}{!}{\includegraphics{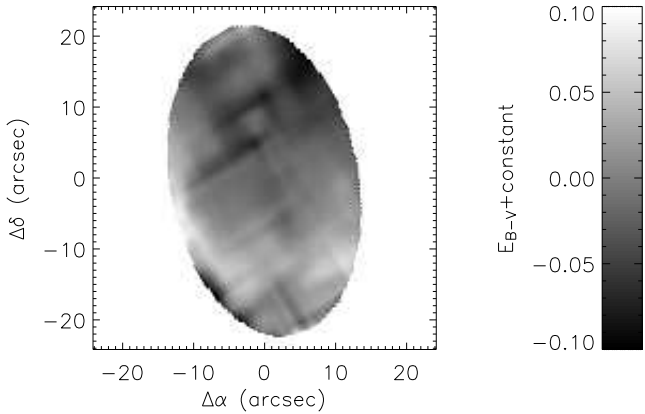}}
\caption{Map of $E_{B-V}$ established with the continuum-free H$\alpha$ and
H$\beta$ images. The average uncertainty of this map is $\approx0.03$. The
computed reddening outside the ellipse cutout is less accurate, so it is not
shown here.}
\label{fig_ebvmap}
\end{figure}
We retrieved several narrow-band images of the ING archive. They were obtained
through filters centered on H$\alpha$, H$\beta$, [O{\sc iii}] $\lambda$5007 and
on the red side of the H$\alpha$ continuum, at the wavelength $\lambda=6384$
{\AA}. The H$\alpha$ filter is also transparent to the [N{\sc ii}]
$\lambda$65(48+83) doublet, but the latter contributes little to the signal of
the images and was neglected.

Using some stars common to all the images, we subtracted the H$\alpha$
continuum from the three other images, in order to keep only the signal of the
nebular emission lines. We then established a reddening map, as shown in Fig.
\ref{fig_ebvmap}. This map was obtained after smoothing the H$\alpha$ and
H$\beta$ images with a 31$\times$31 pixel mask, due to the small S/N ratio of
the latter image. It can be seen that the reddening in the nebula is quite
uniform, so we replaced it with a constant reddening. This simplification
causes an error of only $\sim1$\% in the computation of the total dereddened
H$\beta$ flux. We also established an [O{\sc iii}] $\lambda$5007/H$\alpha$ map,
as shown in Fig. \ref{fig_o3hamap} and discussed in Appendix \ref{app_abk}.

Knowing the position of the optical slit on the nebula and the H$\beta$ flux
measured through it, and assuming a distance of 800 kpc to the nebula
\citep{lsg02}, we calculated the total extinction-corrected H$\beta$ luminosity
of \object{NGC 588}: $L_\beta=1.22\times10^{38}$ erg s$^{-1}$. We tested this
luminosity by using the radio flux density at 1.4 GHz measured by \cite{vgv86}.
The observed radio-to-H$\beta$ ratio is $F_\nu(1.4~{\rm
GHz})/F_\beta=(3.3\pm0.2)\times10^{-14}$ Hz$^{-1}$. We used the emissivity
formulae of \cite{ppb91} for H$\beta$ and of \cite{o74} for the radio
continuum. In the latter, we included the free-free emissions induced by the
H$^+$ and He$^+$ ions. Adopting a temperature of 11000 K and an He$^+$/H$^+$
abundance ratio of 0.083 \cite{vpd88}, we found a theoretical ratio
$F_\nu(1.4~{\rm GHz})/F_\beta=3.66\times10^{-14}$ Hz$^{-1}$, a value higher by
10\% than the observed one. The theoretical value of $F_\nu(1.4~{\rm
GHz})/F_\beta$ depends little on the temperature, and we can assess that the
total dereddened H$\beta$ flux that we found is consistent with the
measurements of \cite{vgv86} to within 10\%. Consequently, for comparison with
the ISO data, we used our value of $L_\beta$ with an associated uncertainty of
10\%.

\subsection{ISO line spectra}
\label{sect_iso}

\begin{figure}
\resizebox{\hsize}{!}{
\hbox{\includegraphics{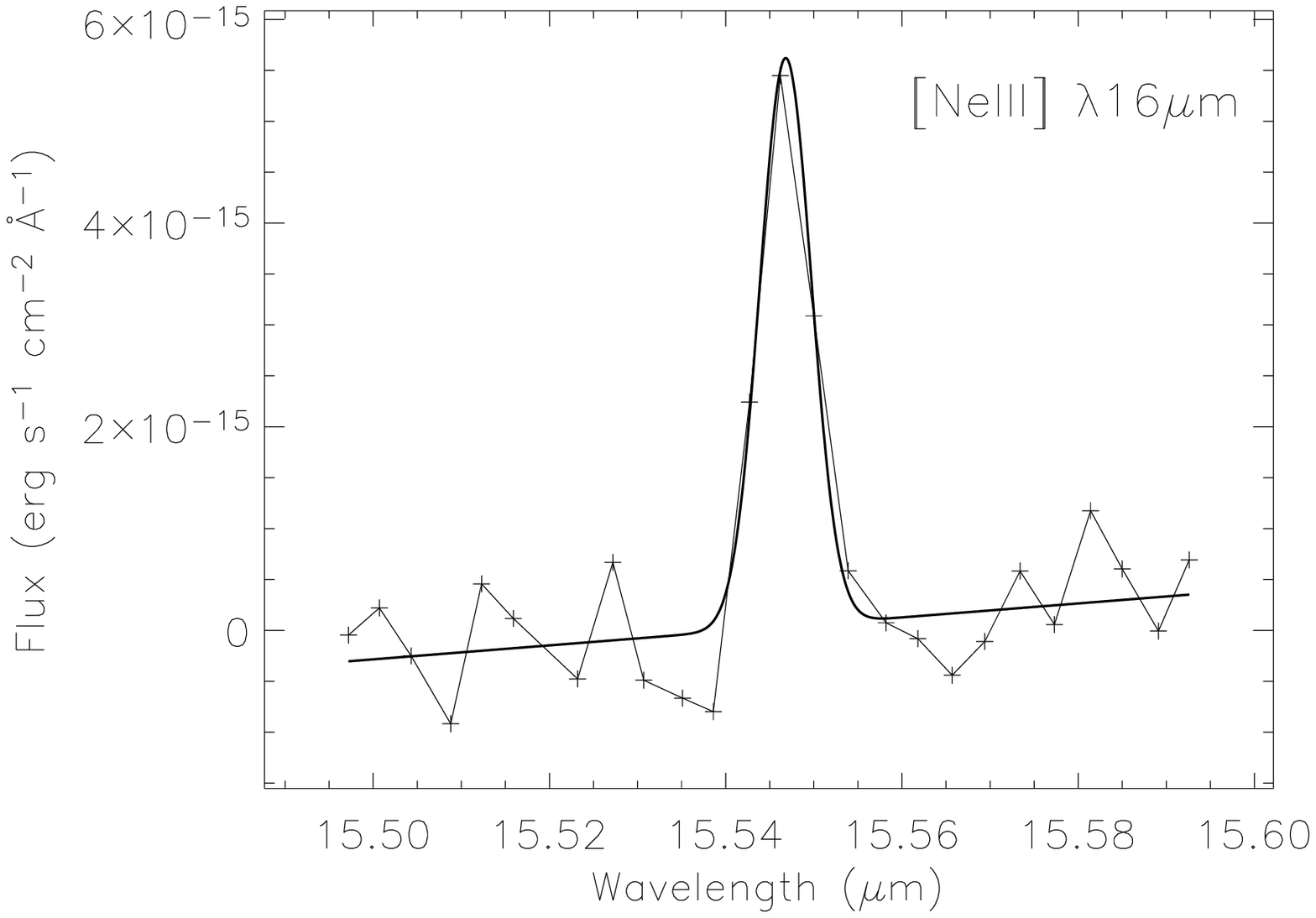}\quad
\includegraphics{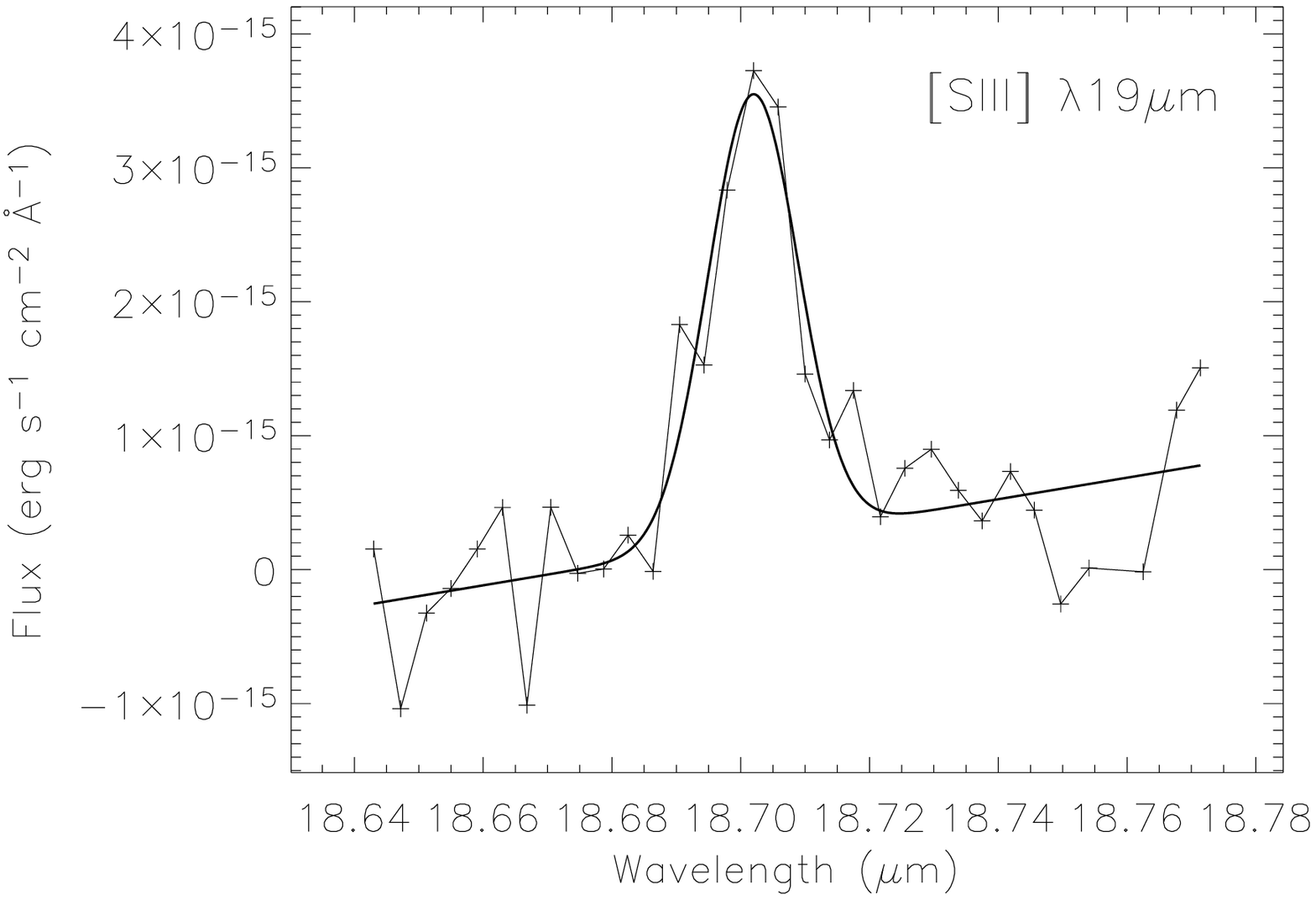}}}
\resizebox{\hsize}{!}{
\hbox{\includegraphics{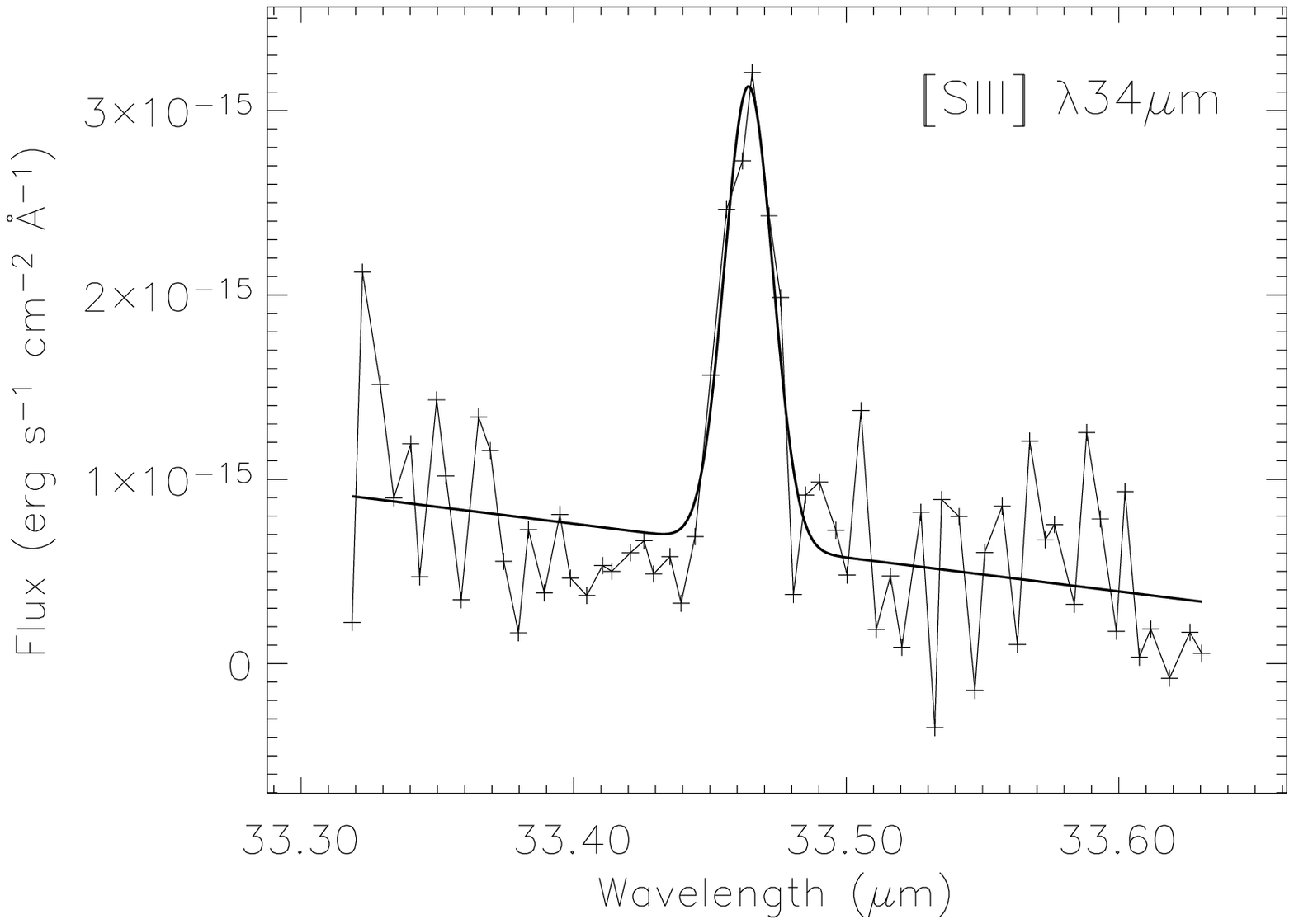}\quad
\includegraphics{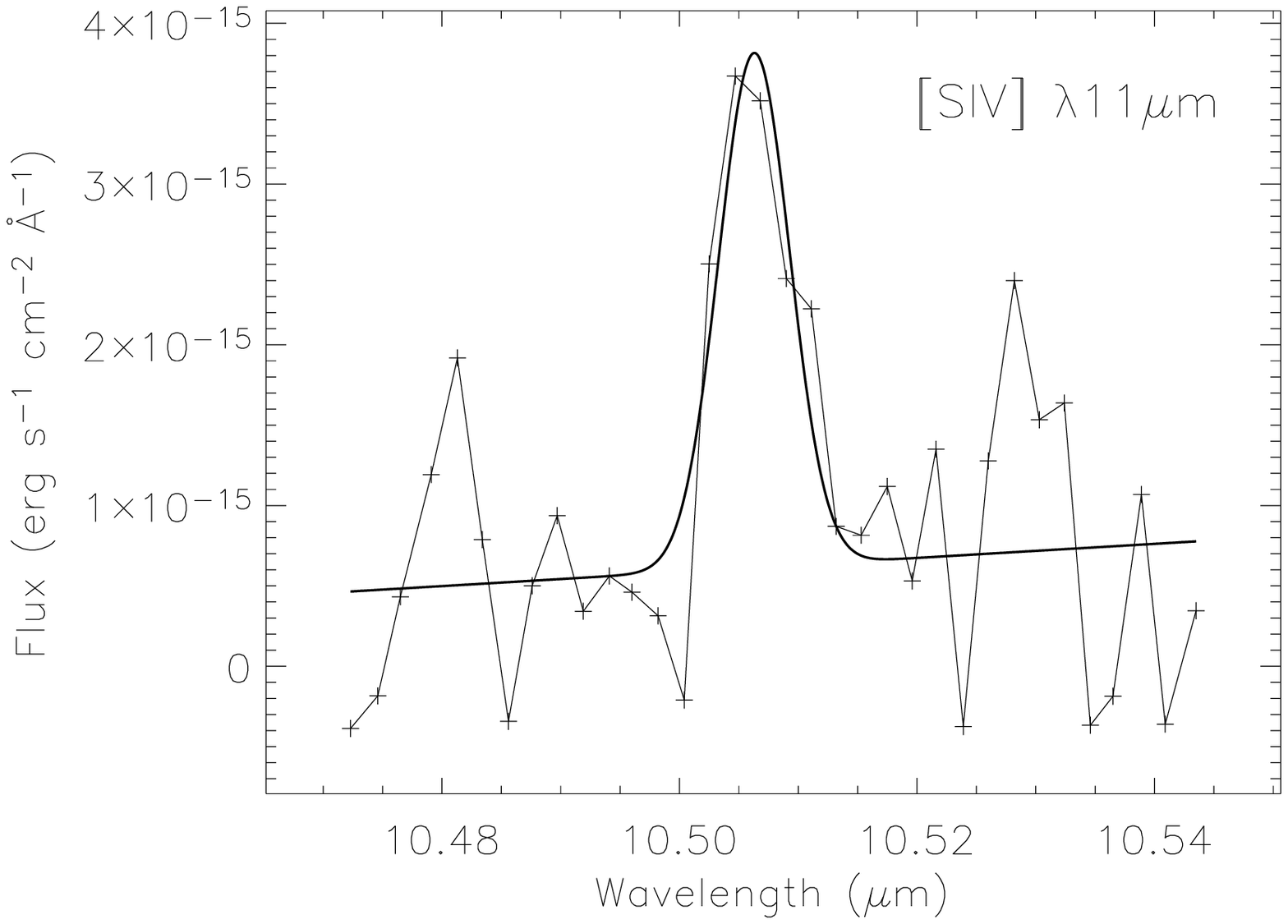}}}
\caption{ISO-SWS detected lines.}
\label{fig_isolines}
\end{figure}

A number of infrared (IR) lines were observed with the Infrared Space
Observatory (ISO), in the ranges of both the Large- and Small-Wavelength
Spectrometers (LWS and SWS, respectively). We retreived 3 data sets from LWS
observations and 1 from SWS observations (Table \ref{tab_obs}).

The LWS data sets 80800268 and 81601776 were analyzed by \cite{hhv03}, who
found a value of $3.51\times10^{-12}$ erg s$^{-1}$ cm$^{-2}$ for the [O{\sc
iii}] $\lambda$88$\mu$m flux. Given the important constraint provided by this
line in our study, we performed our own measurements on the three available
data sets. For this, we used the ISAP\footnote{The ISO Spectral Analysis
Package (ISAP) is a joint development by the LWS and SWS Instrument Teams and
Data Centers. Contributing institutes are CESR, IAS, IPAC, MPE, RAL and SRON.}
facility. For each spectrometer and each detector (\#4 and \#5 for LWS01, \#5
for LWS02), we obtained one spectrum by selecting the scans to combine and by
removing bad pixels. Even when processed, the LWS02 contained spurious
features, so it was rejected. The measurements of the [O{\sc iii}]
$\lambda$88$\mu$m flux on the four LWS01 spectra gave compatible values.
Combining them, we find a flux of  $(3.41\pm0.13)\times10^{-12}$ erg s$^{-1}$
cm$^{-2}$, in agreement with the value given by \cite{hhv03}. In the following
we adopted our value. Accounting for the estimates given in the LWS Handbook
\citep{gsh03} on the repeatability of the LWS01 observations and the absolute
photometric accuracy of this spectrometer, we estimate that the overall
uncertainty in the [O{\sc iii}] $\lambda$88$\mu$m flux is 20\%. Since the LWS
aperture ($84''$) is much larger than the nebula, no aperture correction was
needed for the [O{\sc iii}] $\lambda$88$\mu$m flux.

The SWS spectra were processed with the ISAP facility in the same way as the
LWS data. We measured the fluxes of the lines [Ne{\sc iii}] $\lambda$16$\mu$m,
[S{\sc iii}] $\lambda$19$\mu$m, [S{\sc iii}] $\lambda$33$\mu$m, and [S{\sc iv}]
$\lambda$11$\mu$m (Fig. \ref{fig_isolines}). The SWS apertures range from
$14''\times20''$ to $33''\times20''$ and are therefore smaller than the nebula.
To estimate the total nebular fluxes in each of these lines, we assumed that
their surface brightness distribution is identical to the one observed in the
H$\alpha$ image. This assumption is not entirely correct because of ionization
stratification of the nebula. We tested it with a few photoionization models,
using various ionizing spectra and density distributions, and found that it
leads to an error of $\approx10$\% at most. Considering Table 5.3 of the SWS
Handbook \citep{lks03}, we adopted an instrumental uncertainty of 20\% in
addition to the dispersions in the line fits and the error in the aperture
corrections.

The aperture-corrected intensities of the ISO detected lines are reported in
Table \ref{tab_intens}, together with their non-instrumental uncertainties and
the aperture correction factors. They are given relative to the total H$\beta$
flux of the nebula (Sect. \ref{sect_ing}).

\section{Empirical diagnostics}
\label{sect_empir}

\subsection{Density and temperature}
\begin{figure}
\resizebox{\hsize}{!}{\includegraphics{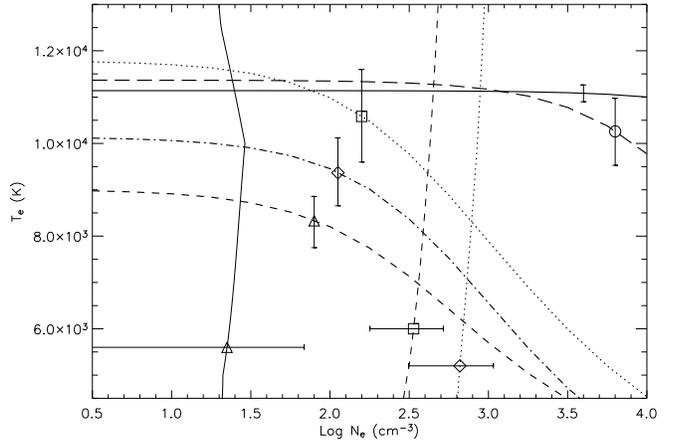}}
\caption{$(T_e,\log N_e)$ diagram for the following ratios. Temperature
diagnostics (with vertical error bars): [O{\sc iii}]
$\lambda$4363/$\lambda$5007 (full line, no symbol), [O{\sc ii}]
$\lambda$73(20+30)/$\lambda$372(6+9) (dotted line, square), [S{\sc ii}]
$\lambda$40(69+76)/$\lambda$67(16+31) (dot-dashed line, diamond), [N{\sc ii}]
$\lambda$5755/$\lambda$6583 (long-dashed line, circle), [O{\sc iii}]
$\lambda$5007/$\lambda$88$\mu$m (short-dashed line, triangle). Density
diagnostics (with horizontal error bars): [S{\sc ii}]
$\lambda$6731/$\lambda$6716 (full line, triangle), [O{\sc ii}]
$\lambda$3726/$\lambda$3729 (short-dashed line, square), [S{\sc iii}]
$\lambda$19$\mu$m/$\lambda$34$\mu$m (dotted line, diamond). The error bars
shown here are representative of the whole $\log N_e$/$T_e$ ranges.}
\label{fig_nete}
\end{figure}

A series of temperature and density diagnostics are available from our observed
line flux ratios. We established a $(T_e,\log N_e)$ curve for five temperature
and three density diagnostics, as seen in Fig. \ref{fig_nete}. In this diagram,
a 7\% uncertainty was set to the [S{\sc ii}] $\lambda$6731/$\lambda$6716 ratio,
in order to account for the uncertainties in the collisional strengths of this
ion. Likewise, the [O{\sc ii}] $\lambda$3726/$\lambda$3729 ratio uncertainty
was set to 15\%.

\begin{figure}
\resizebox{\hsize}{!}{\includegraphics{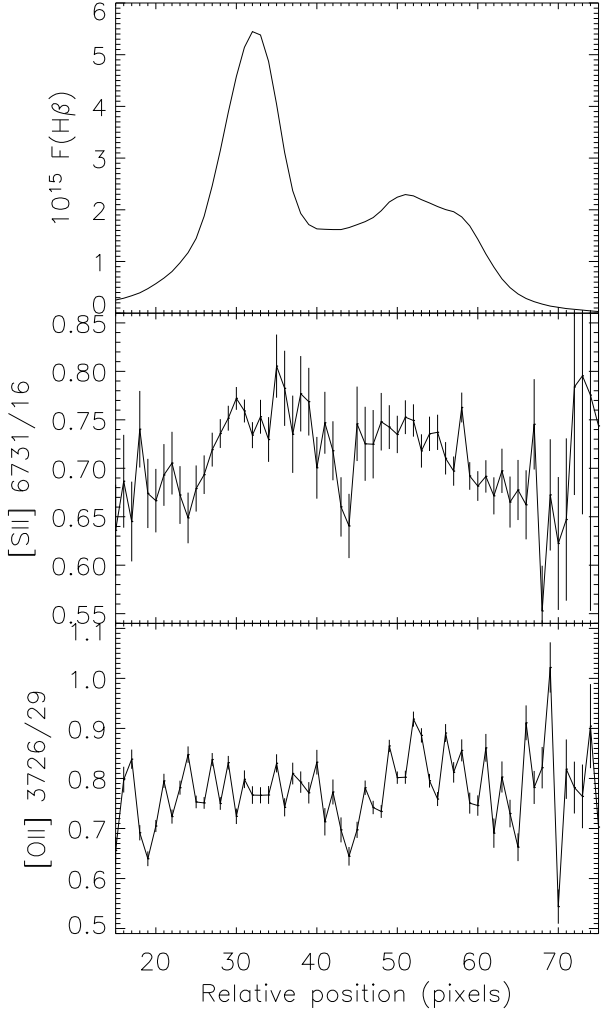}}
\caption{[S{\sc ii}] $\lambda$6731/$\lambda$6716 and [O{\sc ii}]
$\lambda$3726/$\lambda$3729 ratio profiles along the slit. The H$\beta$ flux
profile is shown in the upper panel.}
\label{fig_s2o2prf}
\end{figure}

At temperatures close to 10000 K, a simple least-square fit of the [O{\sc ii}]
$\lambda$3726/$\lambda$3729, [S{\sc iii}] $\lambda$19$\mu$m/$\lambda$34$\mu$m
and [S{\sc iii}] $\lambda$19$\mu$m/$\lambda$34$\mu$m line ratios with a single
electron density yields $N_e=70\pm60$ cm$^{-3}$. However, while the [O{\sc ii}]
and [S{\sc iii}] ratios indicate very similar densities, the [S{\sc ii}] ratio
indicates a smaller one. This is not necessarily inconsistent, since [S{\sc
ii}] lines are emitted further from the ionizing source than [S{\sc iii}] and
[O{\sc ii}] lines, so that the electron density indicated by the [S{\sc ii}]
doublet is expected to be smaller in case of an outward-decreasing gas density.
Figure \ref{fig_s2o2prf} shows the [S{\sc ii}] $\lambda$6731/$\lambda$6716 and
[O{\sc ii}] $\lambda$3726/$\lambda$3729 ratio profiles along the slit with
respect to the H$\beta$ flux distribution. Both ratios are increasing functions
of the electron density. While the [O{\sc ii}] profile is nearly constant, the
[S{\sc ii}] one suggests a decrease in the outskirts of the nebula. However, in
the low-density regime we observe here, interpretation of the [S{\sc ii}] and
[O{\sc ii}] profiles is not fully conclusive. The H$\beta$ flux is another
density indicator and we use it in Sect. \ref{sect_hbdistr}.

\begin{table}
\caption{Empirical temperatures derived from five line ratios}
\label{tab_empte}
\center
\begin{tabular}{ll}
\hline
\hline
Line ratio & Temperature (K) \\
\hline
[O{\sc iii}] $\lambda$4363/$\lambda$5007 & $11140\pm180$ \\
{[O{\sc ii}]} $\lambda\lambda$7320+30/$\lambda$3727 & $11210\pm1180$ \\
{[S{\sc ii}]} $\lambda\lambda$4069+76/$\lambda\lambda$6716+31 & $9860\pm940$ \\
{[N{\sc ii}]} $\lambda$5755/$\lambda$6583 & $11360\pm840$ \\
{[O{\sc iii}]} $\lambda$5007/$\lambda$88$\mu$m & $8420\pm690$ \\
\end{tabular}
\end{table}

Taking an electron density of $70$ cm$^{-3}$, we derived empirical temperatures
from the five corresponding diagnostics. The values are listed in Table
\ref{tab_empte}. As seen, the four diagnostics using only optical lines,
$T_{\rm opt}($O{\sc iii}$)$, $T($O{\sc ii}$)$, $T($S{\sc ii}$)$, and $T($N{\sc
ii}$)$, are in relatively good agreement with a mean temperature of
$\approx11000$ K, $T_{\rm opt}($O{\sc iii}$)$ being by far the most
constraining of the four diagnostics. However, the [O{\sc iii}]
$\lambda$5007/$\lambda$88$\mu$m ratio gives a temperature $T_{\rm IR}($O{\sc
iii}$)$ that is lower by $\approx3000$ K than $T_{\rm opt}($O{\sc iii}$)$. This
discrepancy cannot be attributed to the error bars alone. Indeed, to yield a
temperature of 11000 K, the [O{\sc iii}] $\lambda$5007/$\lambda$88$\mu$m ratio
would have to be 3.6 instead of 1.6. Such a large discrepancy cannot be
attributed to errors of measurement and/or calibration. Instead, it suggests
the presence of large temperature inhomogeneities in the nebula.

The discrepancy between the optical temperature diagnostics and $T_{\rm
IR}($O{\sc iii}$)$ might arise from two phenomena: (i) the optical slit samples
a peculiar zone of plasma hotter than the overall gas of the nebula, making the
temperature diagnostics, representative of this zone higher than $T_{\rm
IR}($O{\sc iii}$)$ which is measured on the whole nebula; or (ii) temperature
variations exist throughout the entire nebula. In the latter case, for each of
the optical temperature diagnostics, the more temperature-sensitive line (e.g.,
[O{\sc iii}] $\lambda$4363) significantly weights the hotter plasma zones,
biasing the temperature measure toward a higher than average value. This bias
also affects $T_{\rm IR}($O{\sc iii}$)$ but to a lower degree, hence making it
smaller than the temperatures derived from optical line ratios.

Assuming that the discrepancy between $T_{\rm opt}($O{\sc iii}$)$ and $T_{\rm
IR}($O{\sc iii}$)$ is not a sampling effect, we can estimate the average $T_0$
and the mean square fluctuations $T_{0}^{2}\,t^2$ of the temperature in the
O$^{++}$ zone, both weighted by $N_e\,N({\rm O}^{++})$. By expanding to order 2
the sensivity of the lines to the temperature and assuming a density of 70
cm$^{-3}$, we found $T_0=8110\pm660$ K and $t^2=0.087\pm0.008$, i.e. the
temperature undergoes fluctuations of about 30\% RMS. This amplitude of
fluctuations is larger than those generally found in H{\sc ii} regions
\citep[typical values of $t^2$ are 0.02--0.05: ][]{ept02}.

\begin{table}
\caption{Empirical abundance ratios of various elements, with the ICFs
specified}
\label{tab_ab}
\center
\begin{tabular}{lll}
\hline
\hline
Ratio & Value & ICF \\
\hline
\hline
\multicolumn{3}{l}{Our work} \\
\hline
He/H & $(8.96\pm0.42)\times10^{-2}$ & He=He$^+$+He$^{++}$ \\
O/H & $(1.45\pm0.27)\times10^{-4}$ & O=O$^+$+O$^{++}$ \\
N/O & $(4.12\pm0.95)\times10^{-2}$ & N/O=N$^+$/O$^+$ \\
Ne/O & $(2.57\pm0.14)\times10^{-1}$ & Ne/O=Ne$^{++}$/O$^{++}$ \\
\hline
\hline
\multicolumn{3}{l}{\cite{vpd88}} \\
\hline
He/H & $8.3\times10^{-2}$ & \\
O/H & $(2.0\pm0.3)\times10^{-4}$ & \\
N/O & $(3.0\pm0.5)\times10^{-2}$ & \\
Ne/O & $(2.5\pm0.5)\times10^{-1}$ & 
\end{tabular}
\end{table}

\subsection{Abundances}
We computed a few empirical elemental abundance ratios, using $N_e=70$
cm$^{-3}$ and the O$^+$ and O$^{++}$ optical temperature diagnostics. The ionic
abundances were derived from lines measured in the optical spectra. The values
and the ionization correction factors (ICFs) used are summarized in Table
\ref{tab_ab}. The error bars account for the uncertainties in the adopted
density, in the empirical temperatures and in the line fluxes. The abundances
are comparable to the ones inferred by \cite{vpd88}. We found the O/H value to
be 30\% of the solar one as given by \cite{l03}. At this stage, we did not
compute the abundances of elements such as S or Ar, due to the uncertainties in
their ICFs.

The empirical elemental abundance ratios presented here served as a starting
point for our modeling of the nebula.

\section{General strategy for model fitting}
\label{sect_mdl}

\begin{table*}
\caption{Observational values of the line ratios used to constrain the
photoionization models and model results in the form
(model$-$observation)/(tolerance). The models are: [FS] Full Sphere; [HB]
Hollow Bubble; [DD1] and [DD2] H$\beta$ constrained Density Distribution, 4.2
Myr spectrum; [DD2] H$\beta$ constrained Density Distribution, 3.6 Myr
Spectrum; [DF] Density fluctuations; [DG] Dust Grains. The geometric
corrections for the models of Sect. \ref{sect_hbo2o3} and further sections are
also written.$^{a}$Values are valid for the whole nebula; the other data are
representative of the optical slit.}
\label{tab_diag_models}
\center
\begin{tabular}{llllllllll}
\hline
\hline
Diagnostics & Observed                                                                   & [FS]    & [HB]    & [DD1]   & [DD2]   & [DDS]   & [DF]    & [DG] \\
            {\it Section} & & \ref{sect_fs} & \ref{sect_hb} & \ref{sect_dd12} & \ref{sect_dd12} & \ref{sect_spec36} & \ref{sect_densfluc} & \ref{sect_dust} \\
\hline
\hline
{\bf Density}\\
{}[S{\sc ii}] $\lambda$6731/$\lambda$6716 & 0.72$\pm$7\%                                 & $-0.1$  & $+0.0$  & $-0.1$  & $+0.0$  & $+0.2$  & $+0.1$  & $+0.7$ \\
{}[O{\sc ii}] $\lambda$3726/$\lambda$3729 & 0.78$\pm$15\%                                & $-2.1$  & $-2.1$  & $-2.1$  & $-2.1$  & $-2.0$  & $-2.1$  & $-1.9$ \\
{}[S{\sc iii}] $\lambda$19$\mu$m/$\lambda$34$\mu$m$^{a}$ & 1.45$\pm$37\%                 & $-1.8$  & $-1.8$  & $-1.8$  & $-1.8$  & $-1.8$  & $-1.8$  & $-1.8$ \\
\hline
{\bf Temperature}\\
{}[O{\sc iii}] $\lambda$4363/$\lambda$5007 & 0.0091$\pm$5\%                              & $-5.5$  & $-4.9$  & $-5.4$  & $-1.6$  & $+0.2$  & $-1.6$  & $+4.5$ \\
{}[O{\sc ii}] $\lambda$73(20+30)/$\lambda$372(6+9) & 0.0218$\pm$17\%                     & $-1.0$  & $-1.1$  & $-1.0$  & $-0.4$  & $-0.2$  & $-0.4$  & $-0.2$ \\
{}[S{\sc ii}] $\lambda$40(69+76)/$\lambda$67(16+31) & 0.0722$\pm$11\%                    & $-2.3$  & $-2.3$  & $-2.3$  & $-1.7$  & $-1.6$  & $-1.8$  & $-1.9$ \\
{}[N{\sc ii}] $\lambda$5755/$\lambda$6583 & 0.0187$\pm$16\%                              & $-1.2$  & $-1.3$  & $-1.2$  & $-0.3$  & $-0.2$  & $-0.5$  & $-0.6$ \\
{}[O{\sc iii}] $\lambda$5007/$\lambda$88$\mu$m$^a$ & 1.64$\pm$23\%                       & $+2.1$  & $+2.3$  & $+2.4$  & $+3.9$  & $+4.4$  & $+3.3$  & $+5.6$ \\
$T_{\rm opt}($O{\sc iii}$)$ (K) & $11140\pm180$                                          & 10080   & 10210   & 10100   & 10840   & 11850   & 10850   & 11950  \\
$T($O{\sc ii}$)$ (K) & $11210\pm1180$                                                    & 10130  & \hspace{0.5em}9990 & 10160   & 10780   & 11030   & 10740   & 10950  \\
$T($S{\sc ii}$)$ (K) & \hspace{0.5em}$9860\pm940$ & \hspace{0.5em}8060 & \hspace{0.5em}8050 & \hspace{0.5em}8070 & \hspace{0.5em}8490 & \hspace{0.5em}8630 & \hspace{0.5em}8440 & \hspace{0.5em}8390 \\
$T($N{\sc ii}$)$ (K) & $11360\pm840$                                                     & 10380   & 10300   & 10380   & 11080   & 11230   & 10960   & 10840  \\
$T_{\rm IR}($O{\sc iii}$)$ (K)$^a$ & \hspace{0.5em}$8420\pm690$ & \hspace{0.5em}9470 & \hspace{0.5em}9600 & \hspace{0.5em}9640 & 10320 & 10550 & 10080 & 11090 \\
\hline
{\bf Excitation}\\
{}[O{\sc iii}] $\lambda$5007/[O{\sc ii}] $\lambda$(3726+9) & 2.86$\pm$16\%               & $-1.7$  & $-4.9$  & $-0.4$  & $+0.2$  & $-0.2$  & $-0.3$  & $-0.3$ \\
{}[S{\sc iii}] $\lambda$6312/[S{\sc ii}] $\lambda$(6716+31) & 0.0631$\pm$25\%            & $+2.0$  & $-1.1$  & $+6.3$  & $+9.7$  & $+4.9$  & $+3.2$  & $+1.5$ \\
{}[Ar{\sc iv}] $\lambda$(4711+41)/[Ar{\sc iii}] $\lambda$7135 & 0.074$\pm$15\%           & $-3.9$  & $-6.4$  & $-3.9$  & $-3.4$  & $-1.2$  & $-0.3$  & $+2.9$ \\
{}[S{\sc iv}] $\lambda$11$\mu$m/[S{\sc iii}] $\lambda$(19+34)$\mu$m$^a$ & 0.270$\pm$37\% & $-1.0$  & $-2.4$  & $-0.9$  & $-0.9$  & $-1.3$  & $-0.8$  & $+0.1$ \\
\hline
{\bf Abundances}\\
{}He{\sc i} $\lambda$5876/H$\beta$ & 0.1193$\pm$5\%                                      & $-0.7$  & $-1.2$  & $-0.1$  & $+0.0$  & $-0.1$  & $+0.0$  & $-0.3$ \\
{}[O{\sc iii}] $\lambda$5007/H$\beta$ & 4.05$\pm$4\%                                     & $-7.1$  & $-13.3$ & $-5.9$  & $-0.8$  & $+0.9$  & $-1.5$  & $+2.2$ \\
{}[N{\sc ii}] $\lambda$6563/[O{\sc ii}] $\lambda$(3726+9) & 0.075$\pm$16\%               & $-0.1$  & $+0.0$  & $-0.4$  & $-1.1$  & $-1.0$  & $-0.6$  & $-0.3$ \\
{}[Ne{\sc iii}] $\lambda$3869/[O{\sc iii}] $\lambda$5007 & 0.0911$\pm$5\%                & $-4.3$  & $-6.0$  & $-4.7$  & $-4.0$  & $-0.5$  & $-0.2$  & $+1.0$ \\
{}[Ar{\sc iii}] $\lambda$7135/H$\beta$ & 0.0946$\pm$5\%                                  & $+1.7$  & $+1.7$  & $+2.7$  & $+6.9$  & $+7.9$  & $+5.6$  & $+7.2$ \\
{}[O{\sc iii}] $\lambda$88$\mu$m/H$\beta^a$ & 2.21$\pm$25\%                              & $-2.2$  & $-2.8$  & $-2.1$  & $-2.0$  & $-2.0$  & $-1.9$  & $-2.1$ \\
{}[Ne{\sc iii}] $\lambda$16$\mu$m/[O{\sc iii}] $\lambda$88$\mu$m$^a$ & 0.257$\pm$33\%    & $+0.2$  & $-0.1$  & $+0.3$  & $+0.3$  & $+1.0$  & $+0.9$  & $+1.5$ \\
\hline
{\bf Other}\\
{}[O{\sc i}] $\lambda$6300/H$\beta$ & 0.02$\pm$50\%                                      & $-1.6$  & $-1.0$  & $-1.9$  & $-2.0$  & $-1.8$  & $-1.7$  & $-1.4$ \\
{}[O{\sc iii}] $\lambda$5007/H$\beta$$^a$ & 3.52$\pm$6\%                                 & $-5.4$  & $-9.5$  & $-4.6$  & $-0.9$  & $+0.8$  & $-1.1$  & $+1.8$ \\
\hline
\hline
{\bf Geometric corrections}\\
$A$ &                                                                                    &         &         & $1.07$  & $1.07$  & $1.40$  & $1.75$  & $1.35$ \\
$K$ &                                                                                    &         &         & $1.48$  & $1.48$  & $0.52$  & $0.52$  & $0.12$ \\
$B$ &                                                                                    &         &         & $0.58$  & $0.58$  & $1.45$  & $3.31$  & $1.15$
\end{tabular}
\end{table*}

To model the nebula, we used the code PHOTO \citep{s90} with the update of
atomic data listed in \cite{s05}. The abundance ratios Mg/O, Si/O, S/O, Cl/O,
Ar/O, Fe/O and C/N were set to the solar values quoted from \cite{l03}. The
input ionizing spectra were those inferred by \cite{jpc04} in their
star-by-star analysis of the cluster, for an age ranging from 3.6 to 4.4 Myr.

The optical slit that we used is thin compared to the nebula, and we can
consider that the line fluxes measured in the optical spectra are
representative of a plane section in the nebula, perpendicular to the sky.
Consequently, those lines were compared to model line intensities integrated in
this plane slice. In the case of the line fluxes representative of the whole
nebula ([O{\sc iii}] $\lambda$5007/H$\beta$ ratio from the images, lines
observed with ISO), the model line intensities were integrated into the whole
gas volume.

Table \ref{tab_diag_models} summarizes the line ratios used to constrain the
models, sorted according to the data they mainly diagnose. The order of the
lines in each ratio was chosen so that the ratio is an increasing function of
the quantity it probes. We considered that satisfactory models should be able
to reproduce all those constraints simultaneously. The observed values are
quoted with their relative tolerances in percentile of the observed values.
For most of the line ratios, those tolerances were defined as the 1$\sigma$
measure uncertainties. The exceptions are the following. We set the tolerances
on [S{\sc ii}] $\lambda$6731/$\lambda$6716 and [O{\sc ii}]
$\lambda$3726/$\lambda$3729 to 7\% and 15\%, respectively, due to the
uncertainties in the collisional strengths of the regarded transitions. We also
increased by 25\% the tolerance for the [S{\sc iii}] $\lambda$6312/[S{\sc ii}]
$\lambda$67(17+31) and [S{\sc iv}] $\lambda$11$\mu$m/[S{\sc iii}]
$\lambda$(19+34)$\mu$m diagnostics, in order to account for the poorly known
dielectronic coefficients of sulfur. In Table \ref{tab_diag_models}, the line
ratios used as temperature diagnostics are followed by the corresponding
empirical temperatures for a density $N_e=70$ cm$^{-3}$.

\section{Simple models}
\label{sect_simple}
\subsection{Full sphere}
\label{sect_fs}
We first modeled the nebula in a very simple way. We assumed it to be a full
sphere of constant density with a radius $R=1.5\times10^{20}$ cm which
approximately fits the orientation-averaged size of the bright annulus that can
be seen in the H$\alpha$ image (Fig. \ref{fig_ha_slits}). The input ionizing
spectrum was the one obtained for an age of 4.2 Myr, the best-fit age from
\cite{jpc04}. We used the total H$\beta$ luminosity of the nebula to constrain
the relation between the filling factor $\epsilon$ and the hydrogen density
$N_{\rm H}$.

We considered several values of $\epsilon$. For $\epsilon=1$, which leads to
the smallest possible density, the model (labelled [FS] in Table
\ref{tab_diag_models}) produced [O{\sc iii}]/[O{\sc ii}], [S{\sc iii}]/[S{\sc
ii}] and [S{\sc iv}]/[S{\sc iii}] ratios that agree satisfactorily with the
data but have too a small value of [Ar{\sc iv}]/[Ar{\sc iii}]. For models with
$\epsilon<1$, all four excitation diagnostic line ratios were underestimated.
However, the main failure of the full sphere model is that it does not
conciliate the temperature diagnostics based on optical lines and the [O{\sc
iii}] optical/IR one. Indeed, it predicts a total amplitude of the variations
of the electronic temperature $T_e$ of $\approx1000$ K only.

\subsection{Hollow bubble}
\label{sect_hb}
The H$\alpha$ image (Fig. \ref{fig_ha_slits}) shows that the nebula is 
better represented by a bubble than by a full sphere. Hence, we modeled the 
nebula as a hollow bubble whose thickness is 20\% of its average radius 
$R=1.5\times10^{20}$ cm. We assumed $\epsilon=1$, leading to a density 
$N_{\rm H}=11$ cm$^{-3}$ (derived from the H$\beta$ luminosity). As seen 
in Table \ref{tab_diag_models}, this model, labelled [HB], fails to 
reproduce $T_{\rm opt}($O{\sc iii}$)$, while it marginally agrees with 
$T_{\rm IR}($O{\sc iii}$)$. It also underestimates the excitation of the 
nebula.

It seems clear that a more complex geometry for the gas is required in the
models. Indeed, a complex density distribution is expected to change the
ionization structure of the nebula, hence affecting the excitation diagnostics.
Such a change would also modify the distribution of the main cooling ions
(e.g., O$^{++}$) and their efficiency to radiate energy away from the nebula.
This may produce spatial temperature variations that are larger than in the
case of a uniform density nebula.

\section{Models with constrained density distribution}
\label{sect_hbo2o3}
We used the well-observed distributions of the H$\beta$, [O{\sc ii}]
$\lambda$372(6+9) and [O{\sc iii}] $\lambda$5007 flux profiles along the
optical slit and the H$\alpha$ and [O{\sc iii}] $\lambda$5007 images to build a
model of the gas density distribution. This model was used in all the
photoionization models that follow.

\subsection{Construction of the density model}
\subsubsection{RMS density assuming spherical symmetry}
\label{sect_hbdistr}
In a nebula, the H$\beta$ flux observed in a column of solid angle $\Omega$ and
coordinates $(x,y)$ is:
\begin{equation}
F_\beta(x,y)~=~\frac{\Omega}{4\pi}\int_{Z_1(x,y)}^{Z_2(x,y)}(\varepsilon_\beta
h\nu_\beta\,\,N_eN_p\,\epsilon)(x,y,z)dz
\end{equation}
where $z$ is the coordinate along the line of sight, $\varepsilon_\beta
h\nu_\beta$, the H$\beta$ emissivity, and $N_e$ and $N_p$, the electron and
proton density. Hence, assuming a given symmetry of the nebula, a model of the
gas density distribution can be derived from the projected $F_\beta(x,y)$
distribution. This is what we did for \object{NGC 588}, from the flux observed
in the CAHA slit and from the H$\alpha$ image.

The H$\alpha$ image suggests that the nebula is formed mainly by a bubble,
possibly containing tenuous gas and surrounded by a halo. It also contains a
bright knot located at the extremity of a filament. This knot and the other
extremity of the filament are covered by the slit. With these considerations,
we derived the following RMS density $N_{\rm H}\,\sqrt{\epsilon}=f(x,z)$ in the
plane covered by the optical slit, from the long-slit profile $F_\beta(x)$:
\begin{itemize}
\item{two ``half-disks'' located on either side of a central line of sight (of
abscissa $x=0$), each composed of a density plateau with $f(x,z)=N_1$ for
$r=\sqrt{x^2+z^2}<R_1$, an annulus with $f(x,z)=N_2>N_1$ for $R_1\leq r<R_2$,
and a fading halo with $f(x,z)=N_3\,(R_2/r)^2$ for $R_2\leq r<R_3$}
\item{two Gaussian round ``blobs'' on either side of the central line of sight,
with the form $f(x,z)=N_G\exp(-((x-x_G)^2+(z-z_G)^2)/(2\sigma_{G}^{2}))$}
\end{itemize}

\begin{figure}
\resizebox{\hsize}{!}{\includegraphics{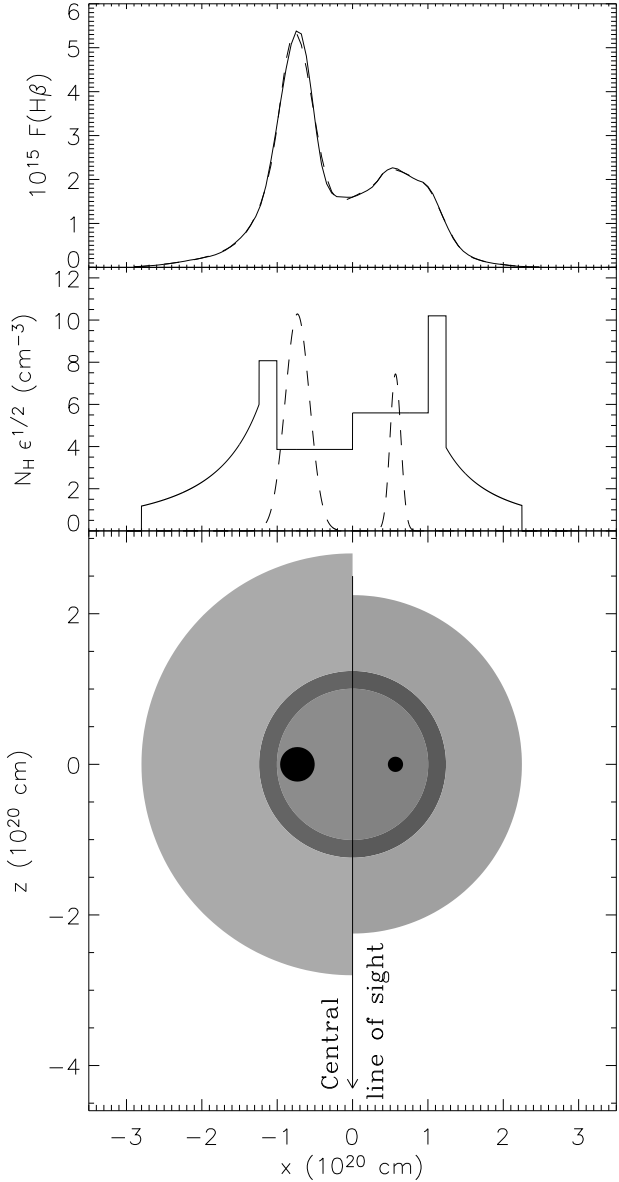}}
\caption{Upper panel: H$\beta$ long-slit profile (full line: observed; dashed
line: model). Middle panel: model of the $N_{\rm H}\,\sqrt{\epsilon}$
distribution (full line: gas surrounding the source; dashed lines: Gaussian
blobs). In this panel, the value of $N_{\rm H}\,\sqrt{\epsilon}$ for the two
Gaussian blobs was divided by 4, for better visibility. Lower panel: schematic
view of the model density profile in the plane section of the optical slit. The
$x$ axis is parallel to the slit, while the $z$ axis is parallel to the line of
sight. The two blobs (represented by black dots) may actually be located at any
position along $z$.}
\label{fig_hbslit}
\end{figure}

This model is presented in Fig. \ref{fig_hbslit}, where the good agreement
between the observed $F_\beta(x)$ profile and the one predicted from the
density model can be appraised. In the model that we fitted, the values of
$N_1$, $N_2$, $N_3$ and $R_3$ depend on the considered side of the central line
of sight ($x<0$ or $x>0$). We simplified this by adopting average values of
those parameters in what follows; these values are listed in the 2nd column of
Table \ref{tab_dhparms}. We also neglected the smaller Gaussian blob, because
of its small contribution to the nebular flux (3\% of the H$\beta$ flux in the
slit). Furthermore, we adopted a distance of $4\times10^{19}$ cm between the
bright blob and the stellar source, equal to the projected distance between
this blob and the main condensation of the stellar cluster. In what follows, we
will call {\it S} the gas distribution that symmetrically surrounds the source,
and {\it G} the bright Gaussian knot. {\it G} contributes to 30\% of the
H$\beta$ flux seen through the optical slit and cannot be neglected.

\begin{table}
\caption{Parameters of the model density distribution, fitted on the slit
H$\beta$ profile and on the image. The distances are expressed in $10^{20}$ cm,
and the densities in cm$^{-3}$.$^{a}$Projected minor and major
semi-axes.$^{b}$Projected distance to the source.}
\label{tab_dhparms}
\center
\begin{tabular}{lll}
\hline
\hline
Parameter & Slit value & Image value \\
\hline
$R_1$ & 1.01 & 0.97$\times$1.61$^{a}$ \\
$R_2$ & 1.24 & 1.19$\times$1.98$^{a}$ \\
$R_3$ & 2.49 & 2.39$\times$3.99$^{a}$ \\
$N_1$ & 4.81 & 6.13 \\
$N_2$ & 9.13 & 10.9 \\
$N_3$ & 5.09 & 6.04 \\
$R_G$$^{b}$ & 0.40 & 0.40 \\
$\sigma_G$ & 0.17 & 0.17 \\
$N_G$ & 40.3 & 40.3 \\
\end{tabular}
\end{table}

We extended the model of density distribution to the case of the whole nebula,
by using the H$\alpha$ image (Fig. \ref{fig_ha_slits}). For this, we stretched
the spherical distribution {\it S} in the image plane, according to the
elliptical shape of the bubble in the image, while conserving the original
dimensions along the line of sight. We then re-fitted $N_1$, $N_2$ and $N_3$ on
the parts of the H$\alpha$ image unoccupied by the bright central filament. The
values obtained, listed in Table \ref{tab_dhparms}, are slightly different from
the ones obtained for the optical slit. {\it G} contributes to $\approx7$\% of
the total H$\beta$ flux of the nebula.

The knot {\it G} is relatively confined and is not centered on the source.
Since our code only processes spherically symmetric gas distributions, we
modeled {\it G} with a shell of Gaussian density profile, but we integrated the
line fluxes in a way adapted to its non-radial gas distribution.

\subsubsection{Accounting for non-sphericity}
\label{sect_asym}
The model presented above assumes spherical symmetry of its different
components, in contradiction with what stands out of the H$\alpha$ image.
Indeed, the projected images of both {\it S} and {\it G} are elliptical,
showing that their gas distributions are not isotropic. Furthermore, such
asymmetry can exist along the line of sight with respect to the plane of the
sky and it that case, it remains undetectable because of the projection of the
signal received from the nebula.

\begin{figure}
\resizebox{\hsize}{!}{\includegraphics{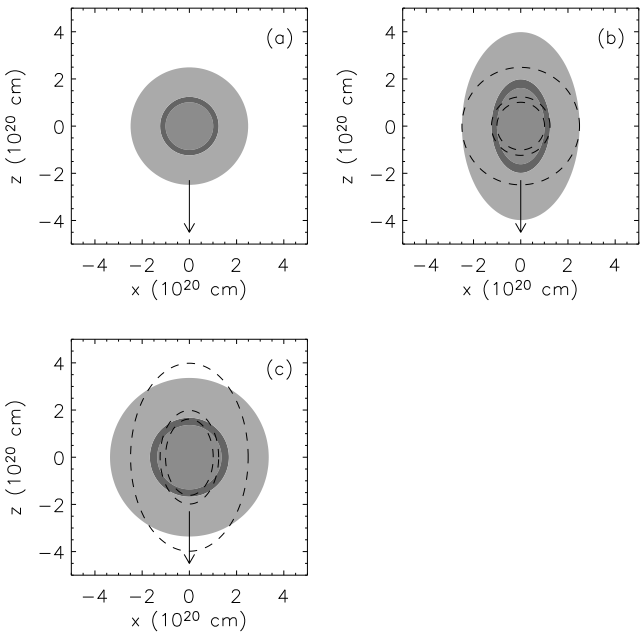}}
\caption{Illustration of the geometrical correction for the gas distribution of
{\it S}. The arrows indicate the direction of the observer.
(a) Model inferred in Sect. \ref{sect_hbdistr}. (b) Possible shape of
the real distribution. (c) Approximation of (b) adopted in photoionization
models in Sect. \ref{sect_hbo2o3}--\ref{sect_further}.}
\label{fig_sim}
\end{figure}

Our photoionization code assumes spherical symmetry, meaning that the accurate
geometry of {\it S} cannot be processed. This is why we decided to model {\it
S} with a spherical gas distribution obtained by homothetic transform of the
three-component distribution $N_{\rm H}\,\epsilon=f(r)$ established in Sect.
\ref{sect_hbdistr}. This transform, illustrated in Fig. \ref{fig_sim}, is meant
to infer sizes to the different structures (e.g., the bubble) that are averages
of the real dimensions of these structures. We used two homothetic factors, $A$
for the lines seen through the optical slit and $B$ for the lines integrated
over the whole nebula. $A$ and $B$ are basically different, because they are
representative of different portions of the nebula.

In Sect. \ref{sect_hbdistr}, we assumed {\it G} to be spherical. Nevertheless,
it may be stretched or squeezed along the line of sight with respect to the
plane of the sky. In this case, the volume of {\it G} is not the one that we
inferred in Sect. \ref{sect_hbdistr}. Since its H$\beta$ flux is fixed, its
density is also different from the one initially inferred. This effect must be
taken into account, because of the importance of the density on the ionization
state of the gas. Hence, we introduced a factor $K$ of stretching {\it G} along
the line of sight with respect to the projected view.

The geometric corrections that we introduced have significant consequences for
the ionization state predicted by the models. In particular, they modify the
[O{\sc iii}] $\lambda$5007/[O{\sc ii}] $\lambda$372(6+9) and [O{\sc iii}]
$\lambda$5007/H$\beta$ ratios in both {\it S} and {\it G}. We used those
ratios to constrain the coefficients $A$, $B$ and $K$, following the procedure
explained in Appendix \ref{app_abk}.

\subsection{Model results}
\label{sect_dd12}
Model [DD1] of Table \ref{tab_diag_models} was established with a filling
factor of $\epsilon=1$. All the parameters other than the density are the same
as those of model [FS] and [HB] (Sect. \ref{sect_simple}). While $T_{\rm
IR}($O{\sc iii}$)$ is marginally reproduced, $T_{\rm opt}($O{\sc iii}$)$ is
still much underestimated.

In order to attempt to fit the optical temperature diagnostics --- and
temporarily ignoring the optical/IR diagnostic --- we modified the input carbon
and sulfur abundances, since these two elements are the most efficient coolants
among those whose abundances were not directly measured. The other parameters,
including $A$, $K$ and $B$, were unchanged. We were able to fit the temperature
diagnostics only with very small carbon and sulfur abundances: even if the
latter are set to 1/10 of the original values, $T_{\rm opt}($O{\sc iii}$)$ and
$T($S{\sc ii}$)$ are almost too small compared to the observed ones (column
[DD2] of Table \ref{tab_diag_models}). Such small abundances are physically
unlikely, so we rejected the model.

\section{Further refining of the models}
\label{sect_further}
\subsection{Using the degree of freedom on the stellar ionizing spectrum}
\label{sect_spec36}
So far, we only used the 4.2 Myr cluster model of \cite{jpc04}, which
corresponds to their best fit model. However, acceptable cluster ages range
from 3.6 to 4.4 Myr. The hardness of the Lyman continuum decreases with
increasing age. Therefore, we used the cluster spectrum corresponding to 3.6
Myr to test whether it is at least able to explain the 11000 K temperature
diagnosed by the optical lines (ignoring again the optical/IR diagnostic).

We discarded the filling factor $\epsilon=1$, because it requires values of $A$
and $B$ that are too large, which is equivalent to assuming that the nebula is
extremely stretched along the line of sight. For $\epsilon=0.1$ and using the
original carbon and sulfur abundances, $T_{\rm opt}($O{\sc iii}$)$ is still
underestimated. However, this issue is resolved if the sulfur abundance is
divided by 2 (model [DDS]), a value that can be considered acceptable.

Although using an ionizing spectrum harder than the one first used improves the
prediction of the optical temperature diagnostics, the large temperature
discrepancy between the latter and $T_{\rm IR}($O{\sc iii}$)$ remains
unexplained.

\subsection{Appending local density fluctuations}
\label{sect_densfluc}
Most GHRs present complex local structures, such as filaments and condensations
\citep[e.g., \object{NGC 604} and \object{30 Dor}: see,
respectively,][]{map04,wmb02}. Such structures may exist in \object{NGC 588}
and remain invisible in our ground-based images due to insufficient angular
resolution. Local structures in the gas are usually modeled by using filling
factors $\epsilon<1$, i.e. assuming that the gas is entirely condensed in
uniform clumps occupying a fraction $\epsilon$ of the available volume.
However, it is more realistic to assume that the density varies continuously,
being higher in the bright structures than in the regions situated between
them.

A continuous density distribution with non-zero density between the bright
structures may increase the amplitude of the spatial temperature variations
since, due to a different distribution of the cooling ions, the temperature in
the diffuse component is expected to be different from that in the filaments.

\subsubsection{Formalism of the fluctuations}
The RMS density of Sect. \ref{sect_hbo2o3} was multiplied by the following
periodic radial function:
\begin{equation}
s(r;H,r_0)~=~\frac{L+(1-L)\,|\cos(\pi\,r/r_0)|^n}{\sqrt{H}}
\end{equation}
where $L=\min(s)/\max(s)$ defines a diffuse gas component between a series of
clumps, whose condensation level is described by the parameter $n$; $H$ is a
normalization factor such that $<f^2>_r=1$. $H$ ensures that the observed
H$\beta$ flux distribution will not be modified within scales larger than $r_0$
since, locally, the H$\beta$ is proportional to $s(r)^2$. We note that
$\max(s^2)=1/H$. Furthermore, $H$ only depends on $L$ and $n$, and the actual
value $r_0$ is only required to be small, in order to simulate small-scale gas
structures. Finally, we will deal not with $L$, but with $\phi=L^2/H$, the
fractional contribution of the diffuse component to the H$\beta$ flux. We
applied the density fluctuations only to {\it S}, and kept the original smooth
density profile for {\it G}.

\subsubsection{Results}
\label{sect_df}
\begin{figure}
\resizebox{\hsize}{!}{\includegraphics{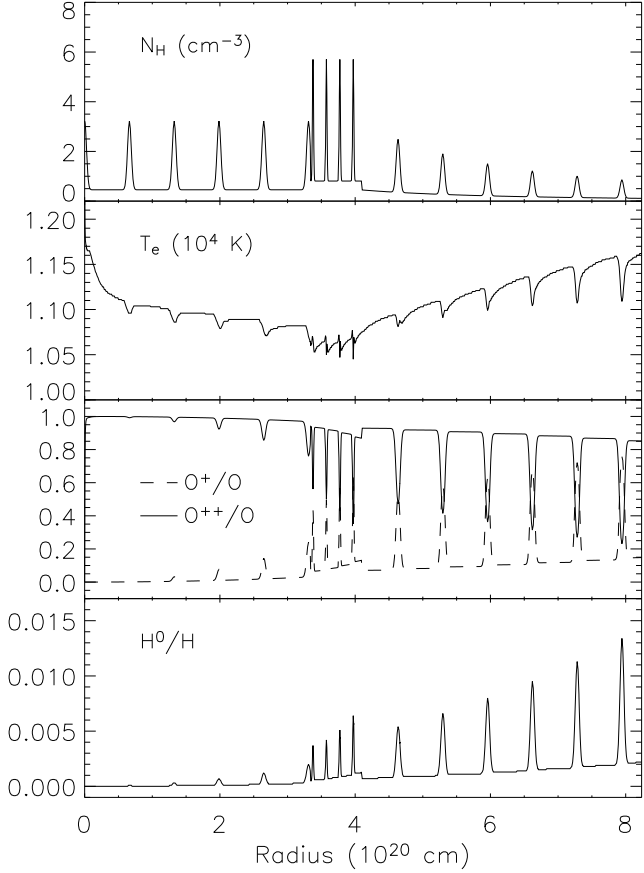}}
\caption{Radial profiles of $N_{\rm H}$, $T_e$, and the relative abundances of
the most coolant ions, in model [DF].}
\label{fig_dtia}
\end{figure}

We computed a few models with various values of $H$ and $\phi$, using the
stellar spectrum for 3.6 Myr and the standard elemental abundances. The
different models gave similar results. The one shown in Table
\ref{tab_diag_models} ([DF]) was obtained for $H=0.1$ and $\phi=0.2$. It can be
seen that the fluctuations of density do not resolve the discrepancy between
$T_{\rm opt}($O{\sc iii}$)$ and $T_{\rm IR}($O{\sc iii}$)$.

Figure \ref{fig_dtia} shows the radial distribution of $N_{\rm H}$, $T_e$,
O$^+$/O, O$^{++}$/O and H$^0$/H in {\it S}. The variations caused by the
density fluctuations to the ionization state induce variations in the cooling
rate of the different chemical elements. In particular, in the outer parts of
the nebula, neutral hydrogen becomes an efficient cooling agent by collisional
excitation of Ly$\alpha$. Unfortunately, the resulting variations of
temperature between the density clumps and the surrounding gaps are smaller
than 500 K.

Close to the source, the temperature fluctuations associated to the clumps are
due in large part to the ionization structure of sulfur. Consequently, we
multiplied the sulfur abundance by 2 and checked the changes in the temperature
distribution. We found that this distribution was modified very little, so we
concluded that density fluctuations are not a solution to the  $T_{\rm
opt}($O{\sc iii}$)$-$T_{\rm IR}($O{\sc iii}$)$ conflict.

\subsection{Models with dust}
\label{sect_dust}
\begin{figure}
\resizebox{\hsize}{!}{\includegraphics{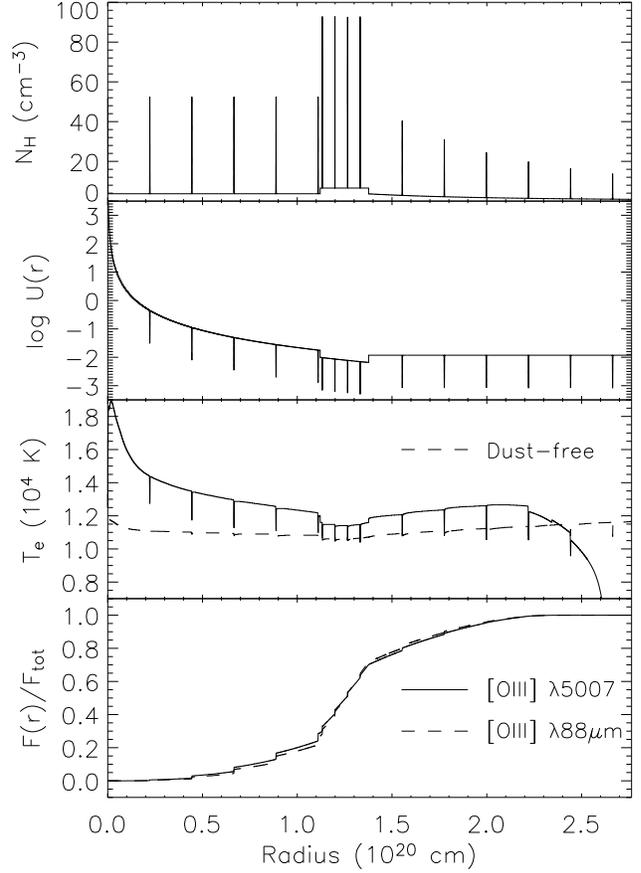}}
\caption{Radial profiles of $N_{\rm H}$, $U$, $T_e$ (three upper panels), and
fractional [O{\sc iii}] $\lambda$5007 and [O{\sc iii}]$ \lambda$88$\mu$m fluxes
integrated through increasing circular apertures (lower panel), in model [DG].}
\label{fig_wdust}
\end{figure}
The effects of dust in the thermal balance of planetary nebulae were studied by
\cite{ss01} with detailed models. They show that dust grains provide additional
heating by photoelectric emission, proportional to the dust-to-hydrogen mass
density ratio $\rho_{\rm d}/\rho_{\rm H}$ and to the local ionization parameter
$U(r)=Q({\rm H}^0)/(4\pi r^2\,N_e\,c)$. Hence, in a nebula where the dust
density and the average ionization parameter are large enough, the decrease in
$U(r)$ with distance $r$ to the source causes a significant negative gradient
of the temperature as a function of $r$. Local density fluctuations also change
the temperature significantly toward lower values in clumps due to a smaller
value of $U(r)$ in them.

We adopted the grain size distribution of \cite{mrn77} for half the total grain
mass and a small grain population (with grain sizes $<0.01$ $\mu$m) for the
other half mass. We limited our study of dust grain effects to the case of the
cluster spectrum for 3.6 Myr, which gives the highest value of $Q({\rm H}^0)$
among our set of cluster spectra. This allowed us to increase $U(r)$ as much as
possible. It also allowed us to maximize the threshold of $\rho_{\rm
d}/\rho_{\rm H}$ above which dust grains would absorb too large a fraction of
the ionizing photons and cause the ionization bound to be closer to the source
than the observed bound. This threshold is $\rho_{\rm d}/\rho_{\rm
H}=5\times10^{-4}$ for the adopted ionizing stellar spectrum.

We adopted $H=0.01$ and $\phi=0.5$ for the density fluctuation function. With
these parameters, the diffuse gas and the clumps contribute equally to the
H$\beta$ flux and $U(r)$, which is inversely proportional to $s(r)$, changes by
a factor as large as $1/\sqrt{\phi H}=14$ between the maxima of the clumps and
the diffuse gas component.

Column [DG] of Table \ref{tab_diag_models} shows the line ratios predicted with
$\rho_{\rm d}/\rho_{\rm H}=5\times10^{-4}$. Even with such a high dust-to-gas
density ratio, we were unable to obtain temperature fluctuations large enough
to explain the conflict between $T_{\rm opt}($O{\sc iii}$)$ and $T_{\rm
IR}($O{\sc iii}$)$ and both temperatures are largely overestimated. Indeed, in
addition to a global increase of temperature, the three effects of dust on the
thermal balance are a significant steepening of the temperature in a small zone
close to the source, which only contributes to $\approx10$\% of the [O{\sc
iii}] $\lambda$5007 and [O{\sc iii}] $\lambda$88$\mu$m fluxes, a drop of the
temperature in the faint outskirts of the nebula and a drop of the temperature
in the clumps by up to $\approx2000$ K, which is insufficient to explain the
temperature discrepancy between the observed diagnostics. This is illustrated
in Fig. \ref{fig_wdust}, where the radial profiles of $N_{\rm H}$, $T_e$, and
$U(r)$ are plotted together with the fractional fluxes of the [O{\sc iii}]
$\lambda$5007 and [O{\sc iii}] $\lambda$88$\mu$m integrated through circular
apertures whose radii are given by the abscissa axis.

To reproduce at least $T_{\rm opt}($O{\sc iii}$)$, $\rho_{\rm d}/\rho_{\rm H}$
would have to be smaller than in model [DG]. In that case, by being smaller
than in the latter, the amplitude of the temperature variations would be even 
further from accounting for the $T_{\rm opt}($O{\sc iii}$)$-$T_{\rm IR}($O{\sc
iii}$)$ discrepancy. Dust grains are therefore an unsatisfactory solution.

\subsection{Spatial distribution of the stars}
\label{sect_specdistr}
It is well known that in a specific point of a nebula, the energy gain provided
by photoionization mainly depends of the hardness of the local Lyman continuum
\citep[e.g.,][]{gst04}, and in particular on the temperature $T_*$ given by:
\begin{equation}
\frac{3}{2}k\,T_*=
\frac{\int_{\nu_0}^{+\infty}(1-\nu_0/\nu)F_\nu\sigma_\nu\,d\nu}
{\int_{\nu_0}^{+\infty}F_\nu\sigma_\nu/(h\nu)\,d\nu}
\end{equation}
where $F_\nu$ is the flux received from the stars, $h\nu_0$ the hydrogen
ionization threshold, and $\sigma_\nu$ the hydrogen ionization cross-section.
In the case of \object{NGC 588}, the ionizing stars are distributed on spatial
scales comparable to the size of the nebula, and one may expect the Lyman
continuum received to vary from point to point, as it is harder close to hotter
stars. In that case, the temperature structure of the gas is sensitive to the
distribution of the stars.

\begin{figure}
\resizebox{\hsize}{!}{\includegraphics{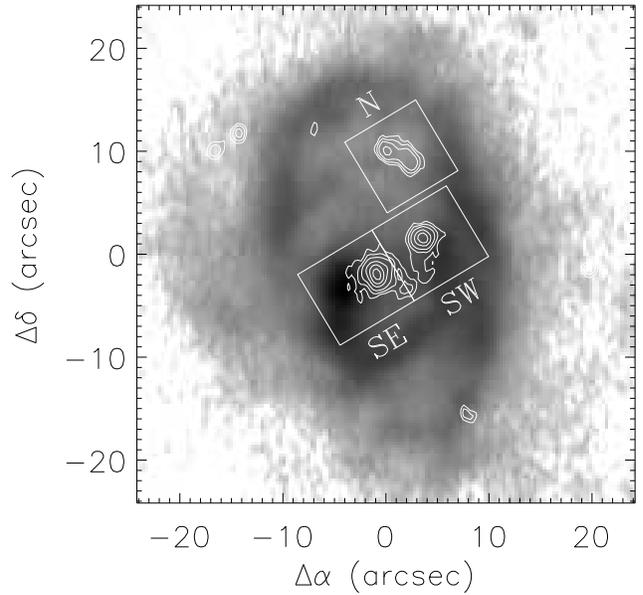}}
\caption{Zones of extraction of spectra in Sect. \ref{sect_specdistr}}
\label{fig_speczones}
\end{figure}

To evaluate the impact of the heterogeneity of the cluster on the temperature
structure of the gas, we divided the cluster into three zones, indicated in
Fig. \ref{fig_speczones}, and extracted the stellar spectrum of each of them,
adopting the age of 3.6 Myr for the cluster. As shown in Table
\ref{tab_speczones}, the three zones individually produce a significant
fraction of the ionizing photons and have different temperatures $T_*$.
Altogether they provide all the ionizing flux of the cluster. The zones SE,
close to the knot {\it G}, and SW, also covered by the optical slit, are
dominated by the two Wolf-Rayet stars (WR) of the cluster, i.e. respectively
stars \#2 and \#1 of \cite{jpc04}. They are the brightest and hottest two stars
of the cluster. Zone N is dominated by a group of three stars, much cooler and
less luminous than the two WR.

\begin{table}
\caption{some spectral properties of the three zones N, SW and SE and of the
whole cluster, assuming an age of 3.6 Myr. $Q_{\rm tot}({\rm H}^0)$ is the
total photon luminosity of the cluster in the Lyman continuum range.}
\label{tab_speczones}
\center
\begin{tabular}{lll}
\hline
\hline
Zone & $Q({\rm H}^0)/Q_{\rm tot}({\rm H}^0)$ & $T_*$ (kK) \\
\hline
N & 25\% & 34 \\
SW & 26\% & 38 \\
SE & 49\% & 48 \\
All & 100\% & 41
\end{tabular}
\end{table}

We re-computed the model [DDS] of Sect. \ref{sect_spec36}, but re-adopted the
original sulfur abundance, and successively replaced the spectrum of the
cluster by the ones of the three zones N, SW, and SE, each rescaled to the
ionizing rate of the whole cluster. The results are shown in Table
\ref{tab_zonetemp}, in terms of empirical temperatures at $N_e=70$ cm$^{-3}$.
While the spectra of N and SW result in very similar temperatures of the gas,
the one of SE yields temperatures higher than the latter by $\approx1500$ K.
Thus, close to the stars of the zone SE, the gas temperature may undergo a
significant excess with respect to the overall nebula. This is the case of {\it
G}, in particular, much closer to SE than to the other parts of the cluster, at
least in the projected sight of the nebula. Hence, we may a priori suspect {\it
G} of being responsible for an increase in $T_{\rm opt}($O{\sc iii}$)$, since
the latter is measured through the optical slit, where {\it G} provides
$\approx$40\% of the [O{\sc iii}] $\lambda$5007 flux. However, by replacing the
model of {\it G} inferred from the total cluster spectrum with the one derived
from the spectrum of WE, we oberved an increase of $T_{\rm opt}($O{\sc iii}$)$
by 300 K only.

The spatial structure of the cluster of \object{NGC 588} turns out not to
provide a satisfactory explanation of the $T_{\rm opt}($O{\sc iii}$)$-$T_{\rm
IR}($O{\sc iii}$)$ conflict that we encountered.

\begin{table}
\caption{Empirical temperatures yielded by the spectra of the zones N, SW and
SE and of the whole cluster.}
\label{tab_zonetemp}
\center
\begin{tabular}{llllll}
\hline
\hline
zone & $T_{\rm opt}($O{\sc iii}$)$ & $T($O{\sc ii}$)$ & $T($S{\sc ii}$)$ &
$T($N{\sc ii}$)$ & $T_{\rm IR}($O{\sc iii}$)$ \\
\hline
N & 9990 & 9860 & 7960 & 10090 & 9580 \\
SW & 10220 & 10080 & 8070 & 10300 & 9760 \\
SE & 11610 & 11450 & 8830 & 11610 & 10880 \\
All & 10840 & 10740 & 8430 & 10920 & 10250
\end{tabular}
\end{table}

\section{Sources of energy other than the ionizing flux}
\label{sect_other}
\subsection{Shocks}
A hypothesis that is often advocated to explain the large temperature
inhomogeneities observed in some GHRs is the supply of energy in the form of
shocks resulting from the winds of massive stars and from supernovae. Those
shocks are expected to raise the temperature abruptly in thin slabs of gas.
However, alteration of the thermal structure induced by shocks in nebulae has
been studied little \citep[see, however, the case of NGC 2363:][]{lcb01}. A
``hot-spot'' model of this phenomenon has been proposed by \cite{bl00}, which
we applied it to \object{NGC 588}.

No supernova remnant was identified within \object{NGC 588} either in the X-ray
survey of \cite{pmh04} or in the radio--optical one of \cite{gdk99}. We
therefore assumed that the kinetic power released by the stellar cluster is
conveyed only by winds. Using the theoretical wind powers implemented in
Starburst99 \citep{lsg99}, we found that the stars of \object{NGC 588} release
a kinetic power $\dot{E}=1.1\times10^{38}$ erg s$^{-1}$, close to the H$\beta$
luminosity of the nebula. On the other hand, the photoionization gain of the
nebula is 15--20 times the H$\beta$ luminosity. As a result, \emph{if we assume
that all the energy of the shocks enhances the collisionally excited lines of
the nebula}, the term $\Gamma_{\rm heat}$ of \cite{bl00} is 0.05--007. Given
the low metallicity of the nebula, this leads to temperature fluctuations below
$t^2=0.01$. This is clearly insufficient to resolve the $T_{\rm opt}($O{\sc
iii}$)$-$T_{\rm IR}($O{\sc iii}$)$ conflict. To reach $t^2=0.087$, the kinetic
power of the stellar winds would have to be at least 5 times larger than that
computed.

\subsection{Electron conduction from hotter plasma layers}
Hot plasma slabs generated by shocks in a nebula are expected to convey heat to
the surrounding gas by electron conduction, creating transition zones between
those slabs and the gas excited by photoionization alone. \cite{mme96}
constructed a model nebula consisting of gas sheets, each of width ${\cal H}$,
separated by thin layers of very hot and tenuous gas. They studied the thermal
structure of the sheets undergoing heat conduction by electrons, and showed
that this structure is able to enhance the [O{\sc iii}] $\lambda$4363 line,
while lines such as H$\beta$ and [O{\sc iii}] $\lambda$5007 are modified
little. Through their Eqs. (9), (10), and (14)--(18), they suggest a relation
between ${\cal H}$, the density $N_p$, the temperature $T_{\rm eq}$ in zones
undergoing photoionization alone, and the temperature diagnostic $T_{\rm
opt}($O{\sc iii}$)$. Adopting $T_{\rm eq}=8000$ K and ignoring Eq. (18), we
found ${\cal H}\sim10^{15}$--$10^{16}$ cm, within the range of densities $N_p$
inferred in Sect. \ref{sect_hbdistr}. This estimate is of course rough, because
it does not accurately take into account the density structure of the nebula
(smooth or with clumps), the ionization state of the gas, and the real value of
$T_{\rm eq}$. However, given the size of the nebula, our estimate of ${\cal H}$
holds the presence of $\sim10^{4}$--$10^{5}$ layers of very hot plasma; even
with higher values of ${\cal H}$, the number of those layers would be much too
large to be realistic. We can therefore reject the electron conduction of heat
as an explanation of the large difference between $T_{\rm opt}($O{\sc iii}$)$
and $T_{\rm IR}($O{\sc iii}$)$.

\section{Uncertainties in O/H and Ne/O}
\label{sect_ohneo}

\begin{table*}
\caption{O$^{++}$/H$^{+}$ and Ne$^{++}$/O$^{++}$ abundance ratios derived from
different diagnostics. The error bars include the uncertainties related to the
temperature diagnostics and to the [O{\sc iii}]/H$\beta$ and [Ne{\sc
iii}]/[O{\sc iii}] line ratios that we used.}
\label{tab_oh_neo}
\center
\begin{tabular}{llll}
\hline
\hline
Set of data & Temperature & Line ratio & Abundance ratio \\
\hline
\hline
\multicolumn{4}{l}{O$^{++}$/H$^{+}$}\\
Optical slit & $T_{\rm opt}($O{\sc iii}$)=1140\pm180$ K &
[O{\sc iii}]$\lambda$5007/H$\beta=4.05\pm0.18$ &
$(1.04\pm0.07)\times10^{-4}$ \\
Integrated & $T_{\rm IR}($O{\sc iii}$)=8420\pm690$ K &
[O{\sc iii}]$\lambda$5007/H$\beta=3.52\pm0.20$ &
$(2.55\pm0.72)\times10^{-4}$ \\
\hline
\hline
\multicolumn{4}{l}{Ne$^{++}$/O$^{++}$}\\
Optical slit & $T_{\rm opt}($O{\sc iii}$)=1140\pm180$ K &
[Ne{\sc iii}] $\lambda3869$/[O{\sc iii}]$\lambda5007=0.0911\pm0.0047$ &
$0.257\pm0.014$ \\
Integrated & $T_{\rm IR}($O{\sc iii}$)=8420\pm690$ K &
[Ne{\sc iii}] $\lambda16\mu$m/[O{\sc iii}]$\lambda88\mu$m$=0.257\pm0.085$ &
$0.164\pm0.059$ \\

\end{tabular}
\end{table*}

Since the thermal balance of \object{NGC 588} is not fully understood, the
estimates of abundance ratios, either empirical or obtained with models, are
uncertain not only through the line ratios that are used to probe them, but
also through the discrepancy that exists between the different temperature
diagnostics. We estimated this uncertainty for the O$^{++}$/H and
Ne$^{++}$/O$^{++}$ abundance ratios, as O$^{++}$ and Ne$^{++}$ are the two ions
for which we have observations in both optical and IR ranges. We compared the
empirical O$^{++}$/H and Ne$^{++}$/O$^{++}$ ratios obtained by using the
long-slit values of [O{\sc iii}] $\lambda$5007/H$\beta$, [Ne{\sc iii}]
$\lambda$3869/[O{\sc iii}] $\lambda$5007, and $T_{\rm opt}($O{\sc iii}$)$ with
those derived from the integrated values of [O{\sc iii}]
$\lambda$5007/H$\beta$, [Ne{\sc iii}] $\lambda$16$\mu$m/[O{\sc iii}]
$\lambda$88$\mu$m, and $T_{\rm IR}($O{\sc iii}$)$.

The results are reported in Table \ref{tab_oh_neo}. It can be seen that the
computed O$^{++}$/H abundance ratio depends a lot on the selected temperature
diagnostic and that the actual value of O$^{++}$/H is known to within $\pm0.2$
dex. Since O$^{++}$ is the dominant ion of O in the nebula, the elemental
abundance ratio O/H is uncertain by a similar amount. Assuming that the
empirical O$^{++}$/O derived from the optical slit data is correct
(O$^{++}$/O=1.39), and depending on the adopted [O{\sc iii}] temperature, O/H
ranges approximately from $1.5\times10^{-4}$ to $3.6\times10^{-4}$. The two
estimates of the Ne$^{++}$/O$^{++}$ ratio, to which Ne/O is expected to be
close, differ by a smaller factor of 1.6.

It is interesting to note that the value of Ne$^{++}$/O$^{++}$ derived from
$T_{\rm IR}($O{\sc iii}$)$ and [Ne{\sc iii}] $\lambda$16$\mu$m/[O{\sc iii}]
$\lambda$88$\mu$m closely matches the solar Ne/O ratio recommended by
\cite{l03}, which is 0.15. This agrees with the usual idea that Ne/O is close
to solar in environments such as GHRs \citep[e.g., ][]{g04}. Furthermore, the
O/H estimate derived from $T_{\rm IR}($O{\sc iii}$)$ and the integrated [O{\sc
iii}] $\lambda$5007/H$\beta$ ratio is found to be close to the average one of
supergiant stars of \object{M33} situated at similar galactocentric distances
\citep[O/H$\approx3\times10^{-4}$: ][]{uhk05}. Hence, the integrated optical/IR
line ratios suggest ``standard'' abundances in the nebula. On the contrary, the
optical spectrum suggests peculiar abundances with a $\sim$50\% O/H depletion
and a Ne/O excess of $\sim$70 \%.

\section{Conclusion}
\label{sect_concl}
This work was motivated by the frequent use of giant H{\sc ii} regions to
determine the chemical composition of the interstellar medium in galaxies. The
usual methods are simple in principle but need to be validated by realistic
models.

We have carried out a comprehensive study of the giant H{\sc ii} region
\object{NGC 588}. We gathered and carefully combined a large number of
observations consisting of spectroscopic data in the UV, visible, and infra-red
domains, as well as of narrow-band images. The most outstanding property of
this H{\sc ii} region is that the temperature derived from the [O{\sc
iii}]$\lambda$5007/$\lambda$88$\mu$m ratio is lower by about 3000~K than the
temperature obtained with four different optical temperature diagnostics. This
suggests either the presence of large temperature inhomogeneities or the
incorrectness of flux measurements. After a careful analysis of the
observational data, we excluded this latter possibility, because it would imply
that the [O{\sc iii}] $\lambda$88$\mu$m line flux has been overestimated by a
factor of 2. We then constructed photoionization models with the aim of
reproducing the observational data, including all the temperature diagnostics.

In a previous paper \citep{jpc04}, we conducted a detailed star-by-star
analysis of the cluster responsible for the ionization of the H{\sc ii} region
and obtained an ionizing spectral energy distribution based on the most
up-to-date stellar atmosphere models. We used this spectral energy distribution
as input to photoionization models tailored to reproduce all the observed
spectral diagnostics. Our procedure takes the different apertures into account
through which the spectroscopic data are obtained. We started with the simplest
density structures, i.e. a homogeneous sphere and then a bubble. Indeed, most
of our understanding of giant H{\sc ii} regions is based on such models. Since
these simple models failed to reproduce all the temperature diagnostics, we
proceeded to more complex representations. Our strategy was to stick as closely
as possible to the observational constraints provided by the spectra and images
and to explore the free parameters left systematically. These include
abundances of unconstrained elements such as carbon, density fluctuations, dust
grains, spatial extent of the ionizing cluster, shock heating and conductive
heating. No satisfactory solution was found. We stress that, although there are
other cases of H{\sc ii} regions for which published photoionization models
were not able to reproduce the observed temperatures, this is the first time
that such a thorough analysis of a giant H{\sc ii} region has been performed
and such a large number of observational constraints used.

The fact that no acceptable model has been found yet has important
consequences. Two explanations can be advanced:
\begin{itemize}
\item{Either the energy balance of this object is not understood. This then
might also apply to giant H{\sc ii} regions in general \citep[][ and references
therein]{ept02}. In that case, abundances derived from optical lines are not as
accurate as usually thought, even in objects where temperature diagnostics are
available.}
\item{Or, the uncertainty in the [O{\sc iii}] $\lambda$88$\mu$m fluxes measured
by ISO is larger than expected. If this is the case, results based on the
analysis of this line, even in other contexts, should be regarded with
caution.}
\end{itemize}
For the time being, in the case of \object{NGC 588}, we can only say that the
O/H ratio probably lies somewhere between $\approx1.5\times10^{-4}$ and
$3.6\times10^{-4}$. However, we found that the O/H and Ne/O ratios derived from
optical/IR integrated data are more consistent with at least some comparison
sources (supergiants in \object{M33} for O/H, the Sun for Ne/O) than those
derived from the optical spectrum and should probably be preferred to the
latter.

Alternatively to temperature fluctuations, abundance inhomogeneities might be a
solution to the conflict between the temperature diagnostics. They have been
advocated in planetary nebulae to explain the inconsistencies between
abundances derived from optical forbidden lines and from recombination lines
\citep{liu03}. Such conflicts are also observed in giant H{\sc ii} regions
\citep[e.g., ][]{ept02}, albeit to a lesser extent. Though abundance
inhomogeneities are theoretically expected not to occur in H{\sc ii} regions
\citep{t96}, the current view is changing \citep{sea05,tp05}. A measurement of
oxygen recombination line intensities in \object{NGC 588} would provide an
important constraint to investigating the presence and significance of
abundance inhomogeneities in this object. In the case where there the latter is
present, one needs to investigate the precise meaning of the chemical
abundances derived with various techniques.

\begin{acknowledgements}      
We are grateful to Ryszard Szczerba for precious help with the ISO data. We
also thank Lise Deharveng, Ariane Lan\c con, Miguel Cervi\~no and Valentina
Luridiana for very useful suggestions. Funding was provided by French CNRS
Programme National GALAXIES, by Spanish grants AYA-2001-3939-C03-01,
AYA-2001-2089, AYA2001-2147-C02-01 and AYA2004-2703, and by the French-Spanish
bi-lateral program PICASSO/Acci\'on Integrada HF2000-0143.

\end{acknowledgements}

\appendix{}
\section{Correction of Balmer lines used for dereddening}
\label{app_dered}
Let us consider a set of hydrogen Balmer lines, whose fluxes are measured at a
series of positions $x$ along a slit. At a given position $x$, the measured
value of the flux of ``H$X$'', $F_X(x)$, differs from its real value
$F_{X}^{\rm real}(x)$ by a measure error $\delta_X(x)=F_X(x)-F_{X}^{\rm
real}(x)$. Furthermore, $F_{X}^{\rm real}(x)$ is related to $F_{\beta}^{\rm
real}(x)$ by the following formula:
\begin{eqnarray}
2.5\,\log F_{X}^{\rm real}(x)&=&2.5\,\log(F_{\beta}^{\rm
real}(x)\,R_X)\nonumber\\
&&+E_{B-V}(x)\,(f_\beta-f_X)
\end{eqnarray}
where $R_X$ is the theoretical H$X$/H$\beta$ intensity ratio, and $f_x$ the
extinction law. By applying this equation to H$\alpha$, we can derive the
following formula, especially for the bright H$\gamma$ and H$\delta$ lines:
\begin{eqnarray}
(f_\beta-f_\alpha)\,\ln \frac{F_{X}^{\rm real}(x)}{R_X}&=&(f_X-f_\alpha)\,\ln
F_{\beta}^{\rm real}(x)\nonumber\\
&&+(f_\beta-f_X)\,\ln \frac{F_{\alpha}^{\rm real}(x)}{R_\alpha}.
\end{eqnarray}

Analogously, we can write the flux of H$X$ predicted from the observed fluxes
of H$\alpha$ and H$\beta$, namely $F_{X}^{\rm pred}(x)$. Due to the measure
errors, $F_{X}^{\rm pred}(x)$ and $F_X(x)$ are discrepant, and if the errors
are small compared to the fluxes, then the discrepancy term
$\Delta_X(x)=F_X(x)-F_{X}^{\rm pred}(x)$ is given by:
\begin{eqnarray}
(f_\beta-f_\alpha)\,\frac{\Delta_X(x)-\delta_X(x)}{F_X(x)}&=&(f_\alpha-f_X)\,
\frac{\delta_\beta(x)}{F_\beta(x)}\nonumber\\
&&+(f_X-f_\beta)\,\frac{\delta_\alpha(x)}{F_\alpha(x)}
\label{eq_errpred}
\end{eqnarray}

We split the flux errors into three terms: spatially invariant residuals
$\epsilon_X$ of the photometric calibration, stellar absorption features
underlying the nebular Balmer lines, and the usual line flux dispersions (such
as photon noise) $\xi_X(x)=0\pm\sigma_X(x)$. The equivalent widths of the
stellar lines are approximately the same for H$\alpha$, H$\beta$, and H$\gamma$
\citep[e.g.,][]{am72}, but may vary spatially if the stellar content is
heterogeneous. Consequently, we described the stellar contribution to the flux
errors as the sum of the contributions of different subsets $i$ of the cluster,
each being associated to an equivalent width $W_i$ and continua $F_{\lambda
X}^{i}(x)$. The sum of the errors $\delta_X(x)$ is:
\begin{eqnarray}
\frac{\delta_X(x)}{F_X(x)}&=&\epsilon_X-\sum_iW_i\,\frac{F_{\lambda
X}^{i}(x)}{F_X(x)}\pm\frac{\sigma_X(x)}{F_X(x)}.
\label{eq_errterms}
\end{eqnarray}
The linear equation \ref{eq_errpred} applies if the relative errors
$\delta_X(x)/F_X(x)$ are small, i.e. if the photometric errors $\epsilon_X$ are
much smaller than 1, if the equivalent widths of the nebular lines with respect
to the stellar continuum are much greater than the ones of the stellar lines,
and if the signal-to-noise ratios (S/N) $F_X(x)/\sigma_X(x)$ are high.

\subsection{Photometric errors}
Let us consider the set of positions $x$ unaffected by stellar absorption
lines, and where the nebular lines were measured with high S/N. Combining
Eqs. \ref{eq_errpred} and \ref{eq_errterms} give:
\begin{eqnarray}
(f_\beta-f_\alpha)\frac{\Delta_X(x)}{F_X(x)}&=&(f_X-f_\beta)\epsilon_\alpha+
(f_\alpha-f_X)\epsilon_\beta\nonumber\\
&&+(f_\beta-f_\alpha)\epsilon_X\pm\tau_X(x)
\end{eqnarray}
with
\begin{eqnarray}
\tau_{X}^{2}(x)&=&
(f_X-f_\beta)^2\left(\frac{\sigma_\alpha(x)}{F_\alpha(x)}\right)^2+
(f_\alpha-f_X)^2\left(\frac{\sigma_\beta(x)}{F_\beta(x)}\right)^2\nonumber\\
&&+(f_\beta-f_\alpha)^2\left(\frac{\sigma_X(x)}{F_X(x)}\right)^2.
\end{eqnarray}
The contribution $\eta_X$ of the photometric residuals to 
$\Delta_X(x)/F_X(x)$ can be estimated with the following formula:
\begin{eqnarray}
\eta_X&=&\frac{\sum_x (\Delta_X(x)/F_X(x))/\tau_{X}^{2}(x)}{\sum_x
1/\tau_{X}^{2}(x)}.
\end{eqnarray}
$\eta_X$ relates $\epsilon_X$ to $\epsilon_\alpha$ and $\epsilon_\beta$:
\begin{eqnarray}
\epsilon_X&=&\eta_X-\frac{f_X-f_\beta}{f_\beta-f_\alpha}\epsilon_\alpha-
\frac{f_\alpha-f_X}{f_\beta-f_\alpha}\epsilon_\beta.
\label{eq_eta}
\end{eqnarray}
Since they are residuals of the fit of the instrumental photometric response, we
can write $\epsilon_X=0\pm\phi_X$. Furthermore, if they are not correlated
between them, we can assume they minimize the error balance $\chi^2=\sum_X
\epsilon_{X}^{2}/\phi_{X}^{2}$, more specifically given by:
\begin{eqnarray}
\chi^2(\epsilon_\alpha,\epsilon_\beta)&=&\sum_X\frac{1}{\phi_{X}^{2}}
\Big(\frac{f_X-f_\beta}{f_\beta-f_\alpha}\epsilon_\alpha+
\frac{f_\alpha-f_X}{f_\beta-f_\alpha}\epsilon_\beta\nonumber\\
&&-\eta_X\Big)^2.
\end{eqnarray}
The estimations and uncertainties of $\epsilon_\alpha$, $\epsilon_\beta$, and
more generally $\epsilon_X$ can be easily derived from this linear $\chi^2$ and
from Eq. \ref{eq_eta}. The correction of the line fluxes for these errors
consists in multiplying them by $1-\epsilon_X$.
\subsection{Underlying stellar absorption lines}
Making use of fluxes corrected for photometric errors and measured with high
signal-no-noise ratio, we can write, for H$\gamma$:
\begin{eqnarray}
(f_\beta-f_\alpha)\frac{\Delta_\gamma(x)}{F_\gamma(x)}&=&-\sum_i W_i\omega_i(x)
\pm\tau_\gamma(x)\\
\omega_i(x)&=&
(f_\gamma-g_\beta)\frac{F_{\lambda\alpha}^{i}(x)}{F_\alpha(x)}
+(f_\alpha-f_\gamma)\frac{F_{\lambda\beta}^{i}(x)}{F_\beta(x)}\nonumber\\
&&+(f_\beta-f_\alpha)\frac{F_{\lambda\gamma}^{i}(x)}{F_\gamma(x)}.
\end{eqnarray}
The estimate of the equivalent widths $W_i$ is given by the minimization of the
following linear $\chi^2$:
\begin{eqnarray}
\chi^2&=&\sum_x\frac{1}{\tau_{\gamma}^{2}(x)}\left(\sum_iW_i\omega_i(x)+
(f_\beta-f_\alpha)\frac{\Delta_\gamma(x)}{F_\gamma(x)}\right)^2.
\end{eqnarray}

\section{Setting the geometric corrections $A$, $B$, and $K$}
\label{app_abk}
\begin{figure}
\resizebox{\hsize}{!}{\includegraphics{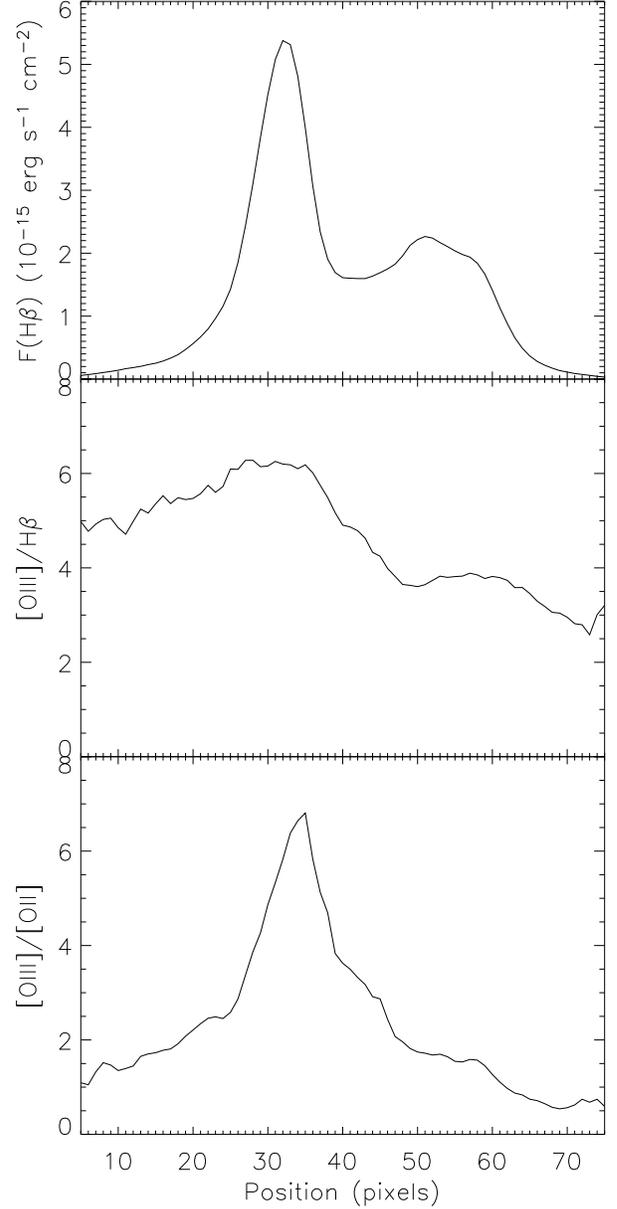}}
\caption{Profiles of H$\beta$ (upper panel), [O{\sc iii}]
$\lambda$5007/H$\beta$ (middle panel), and [O{\sc iii}] $\lambda$5007/[O{\sc
ii}] $\lambda$372(6+9) (lower panel) along the optical slit.}
\label{fig_o3o2o3hb}
\end{figure}

\begin{figure}
\resizebox{\hsize}{!}{\includegraphics{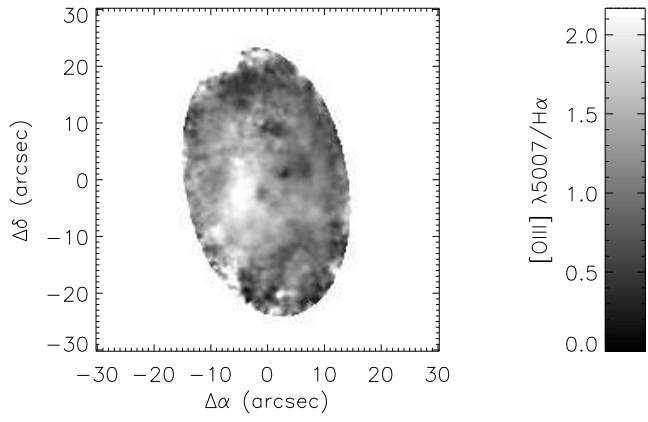}}
\caption{Map of the nebular [O{\sc iii}] $\lambda$5007/H$\alpha$ ratio. The
ratio is too noisy to be shown outside the ellipse cutout.}
\label{fig_o3hamap}
\end{figure}

We used the well-observed long-slit profiles of H$\beta$, [O{\sc iii}]
$\lambda$5007, and [O{\sc ii}] $\lambda$372(6+9), and the [O{\sc iii}]
$\lambda$5007/H$\alpha$ map to constrain the coefficients $A$, $B$ and $K$.
From those data (Fig. \ref{fig_o3o2o3hb} and \ref{fig_o3hamap}), it is evident
that the knot {\it G} is more excited than {\it S}, even though its RMS density
is higher than the latter. This can be easily explained by its short distance
to the stellar source (Sect. \ref{sect_hbdistr}).


Coefficients $A$ and $B$ strongly influence the predicted ionization state of
{\it S}, while $K$ acts on the ionization state of {\it G}. We decided to set
them, for each model, with the following method.

We manually separated the contributions of {\it S} and {\it G} to the long-slit
fluxes of [O{\sc iii}] $\lambda$5007 and [O{\sc ii}] $\lambda$372(6+9), and
derived the values of [O{\sc iii}] $\lambda$5007/[O{\sc ii}] $\lambda$372(6+9)
for both components. We found $F($[O{\sc iii}]$)/F($[O{\sc ii}]$)=2.14\pm0.15$
for {\it S} and $5.68\pm0.65$ for {\it G}.

For each model, we first computed the predicted fluxes of the lines observed
through the optical slit. Unless otherwise mentioned, we set the coefficients
$A$ and $K$ to reproduce the [O{\sc iii}] $\lambda$5007/[O{\sc ii}]
$\lambda$372(6+9) ratios of {\it S} and {\it G}. Once those coefficients were
set, we reported the ratio $R$ between the predicted and the observed value of
[O{\sc iii}] $\lambda$5007/H$\beta$. Then, we calculated the model line fluxes
integrated over the whole nebula. In the latter calculation, coefficient $B$
was chosen so as to reproduce the model/observation ratio $R$, but for the
integrated lines this time.

The geometric transforms performed on the gas density distributions ($f(r)$ for
{\it S}, $g(r)$ for {\it G}) were accompanied with a re-normalization of the
latter to conserve the predicted H$\beta$ fluxes. Indeed, the integrated volume
of gas in {\it S} is proportional to $A^2$ for the optical slit and to $B^3$
for the whole nebula, while the volume of {\it G} is proportional to $K$. Since
the local H$\beta$ emissivity is proportional to the square density, the
transform of $f(r)$ consisted in replacing it by $f(r/A)/A$ (optical slit) or
$f(r/B)/B^{3/2}$ (whole volume), and $g(r)$ was replaced by $g(r)/K^{1/2}$.

\end{document}